\documentclass[reprint,aps,pra,notitlepage,nofootinbib,longbibliography,floatfix,superscriptaddress, preprintnumbers]{revtex4-2}

\usepackage[utf8]{inputenc}
\usepackage{float}
\usepackage{graphicx,nicefrac}  
\usepackage{physics}
\usepackage{esint}
\usepackage{amssymb}  
\usepackage{siunitx}
\usepackage{comment}
\usepackage{mathtools}
\usepackage{amsmath}
\usepackage{amsthm}
\usepackage{wrapfig}
\usepackage{mathrsfs}
\usepackage{enumitem}
\usepackage[colorlinks=true,linkcolor=blue,citecolor=magenta,urlcolor=magenta,plainpages=false,pdfpagelabels]{hyperref}
\usepackage[dvipsnames]{xcolor}
\usepackage{color,colortbl}

\usepackage{bm}
\usepackage[most]{tcolorbox}
\usepackage[normalem]{ulem}

\usepackage{multirow}
\makeatletter 
    
\renewcommand\onecolumngrid{
\do@columngrid{one}{\@ne}%
\def\set@footnotewidth{\onecolumngrid}
\def\footnoterule{\kern-6pt\hrule width 1.5in\kern6pt}%
}

\renewcommand\twocolumngrid{
        \def\footnoterule{
        \dimen@\skip\footins\divide\dimen@\thr@@
        \kern-\dimen@\hrule width.5in\kern\dimen@}
        \do@columngrid{mlt}{\tw@}
}%

\makeatother    

\definecolor{coolblack}{rgb}{0.0, 0.18, 0.39}
\definecolor{darkred}{rgb}{0.5,0,0}
\definecolor{darkgreen}{rgb}{0,0.5,0}
\definecolor{darkblue}{rgb}{0,0,0.5}
\definecolor{lapislazuli}{rgb}{0.15, 0.38, 0.61}
\definecolor{venetianred}{rgb}{0.78, 0.03, 0.08}
\definecolor{bleudefrance}{rgb}{0.19, 0.55, 0.91}
\definecolor{dogwoodrose}{rgb}{0.84, 0.09, 0.41}

\newcommand{\mathcheck}[1]{\color{dogwoodrose}{#1}}
\newcommand{\tcp}[1]{\mathcheck{\tilde{c}_\tp}}

\newcommand{\ci}{\mathrm{i}}
\newcommand{\ee}{\mathrm{e}}

\newcommand{\tv}{\mathrm{v}}
\newcommand{\tx}{\mathrm{x}}
\newcommand{\ty}{\mathrm{y}}
\newcommand{\tk}{\mathrm{k}}

\newcommand{\vx}{\bs{\mathrm{x}}}
\newcommand{\vy}{\bs{\mathrm{y}}}
\newcommand{\vk}{\bs{\mathrm{k}}}
\newcommand{\vq}{\bs{\mathrm{q}}}
\newcommand{\vtv}{\bs{\tv}}

\newcommand{\ai}{a_{\rm i}}
\newcommand{\af}{a_{\rm f}}
\newcommand{\omi}{\omega_\tk^{\rm i}}
\newcommand{\omf}{\omega_\tk^{\rm f}}
\newcommand{\vacin}{\ket{0}_{\rm in}}
\newcommand{\vacout}{\ket{0}_{\rm out}}

\newcommand{\aB}{a_{\text{B}}}

\newcommand{\hb}{\hat{b}}

\newcommand{\bk}{\hat{b}_{\vk}}

\newcommand{\bkin}{\hat{b}_{\vk}^{\rm in}}
\newcommand{\bkout}{\hat{b}_{\vk}^{\rm out}}
\newcommand{\bkdin}{\hat{b}_{\vk}^{\rm in}{}^\dagger}
\newcommand{\bkdout}{\hat{b}_{\vk}^{\rm out}{}^\dagger}
\newcommand{\bmkin}{\hat{b}_{-\vk}^{\rm in}}

\newcommand{\bmkdin}{\hat{b}_{-\vk}^{\rm in}{}^\dagger}

\newcommand{\tho}{\Delta t_{\rm hold}}

\newcommand{\bmk}{\hat{b}_{-\vk}}
\newcommand{\bkd}{\hat{b}^\dagger_{\vk}}
\newcommand{\bmkd}{\hat{b}^\dagger_{-\vk}}

\newcommand{\vna}{\bs{\nabla}}

\newcommand{\lr}[1]{\left(#1\right)}
\newcommand{\lrsq}[1]{\left[#1\right]}

\newcommand{\bs}[1]{\boldsymbol{#1}}
\newcommand{\bS}{\boldsymbol{S}}
\newcommand{\LN}{\text{LN}}

\graphicspath{{./figures/}} 

\begin{document}

\preprint{IPARCOS-UCM-24-059}

\title{Toward the Observation of Entangled Pairs in BEC Analogue Expanding Universes}

\author{Ivan Agullo}
\email{agullo@lsu.edu}
\affiliation{Department of Physics and Astronomy, Louisiana State University, Baton Rouge, LA 70803, U.S.A.}

\author{Adri\`a Delhom}
\email{adria.delhom@gmail.com}
\affiliation{Department of Physics and Astronomy, Louisiana State University, Baton Rouge, LA 70803, U.S.A.}
\affiliation{Departamento de F\'isica Te\'orica and IPARCOS, Facultad de Ciencias F\'isicas, Universidad Complutense de Madrid, Plaza de Ciencias 1, 28040 Madrid, Spain}

\author{Álvaro Parra-López}
\email{alvaparr@ucm.es}
\affiliation{Departamento de F\'isica Te\'orica and IPARCOS, Facultad de Ciencias F\'isicas, Universidad Complutense de Madrid, Plaza de Ciencias 1, 28040 Madrid, Spain}

\begin{abstract}
Pair creation is a fundamental prediction of quantum field theory in curved spacetimes. While classical aspects of this phenomenon have been observed, the experimental confirmation of its quantum origin remains elusive. In this article, we quantify the entanglement produced by pair creation in two dimensional Bose-Einstein Condensate (BEC) analogues of expanding universes and examine the impact of various experimental factors, including decoherence from thermal noise and losses. Our analysis evaluates the feasibility of detecting entanglement in these systems and identifies optimal experimental configurations for achieving this goal. Focusing on the experimental setup detailed in \cite{Viermann:2022wgw}, we demonstrate that entanglement can be observed in these BEC analogues at a significance level of $\sim 2\sigma$ with current capabilities, and at $\gtrsim 3.3\sigma$ with modest improvements. Achieving this would provide unequivocal evidence of the quantum nature of pair creation and validate one of the most iconic predictions of quantum field theory in curved spacetimes.
\end{abstract}
\maketitle


\section{Introduction}

The spontaneous creation of particle pairs triggered by the expansion of the universe was first identified by E.~Schr\"odinger in the late 1930s, referring to it as an \textit{``alarming''} phenomenon \cite{SCHRODINGER1939899}. Schr\"odinger's insights did not attract much attention at the time. This phenomenon was rediscovered by L.~Parker about thirty years later, who, equipped with the modern language of quantum field theory (QFT), was able to firmly establish the phenomenon of spontaneous excitation of entangled pairs by the cosmic expansion \cite{Parker:1968mv, Parker:1969au, Parker:1971pt}. These works laid the foundation for quantum field theory in curved spacetimes (QFTCS), a framework that has led to profound insights and predictions \cite{Hawking74, Hawking:1974rv, Wald:1975kc, Starobinskii:1973hgd, Unruh:1974bw, Fulling:1972md, Davies:1974th, Unruh:1976db}.

As is often the case when discussing the intersection of quantum and gravitational effects, these phenomena are faint and difficult to be observable in the cosmos. Subsequently, Unruh proposed that similar effects might appear in other systems that can be recreated in the laboratory \cite{Unruh:1980cg}. The key observation is that, although many of the predictions of QFTCS were first derived with gravitational backgrounds in mind, Einstein's equations did not play any role, indicating that the fundamental predictions are universal aspects of quantum fields propagating on non-trivial effective geometries, regardless what the physical origin of the geometry is.

Rapid progress in this field has led to outstanding experimental results that confirm classical aspects of QFTCS predictions in experimental analogues of the Hawking effect \cite{philbin_fiber-optical_2008, weinfurtner_measurement_2011,Weinfurtner:2013zfa,Boiron:2014npa,euve_observation_2016, munoz_de_nova_observation_2019, drori_observation_2019, shi_quantum_2023,Falque:2023ctx,Almeida:2022otk}, superradiance \cite{torres_rotational_2017, braidotti_measurement_2022}, pair creation in expanding universes \cite{eckel_rapidly_2018, Banik:2021xjn, Viermann:2022wgw,Svancara:2023yrf}, and the dynamical Casimir effect \cite{jaskula_acoustic_2012}. These observations have been carried out on different physical platforms, confirming the universality of these phenomena \cite{Barcelo:2005fc}.

Significant  efforts are currently focused on detecting entanglement between the created quanta \cite{Busch:2013gna,Busch:2014bza,PhysRevA.89.043808,deNova:2015lva,Robertson:2016evj,Tian:2017cda,Robertson:2017ysi,Jacquet:2020jpj,Agullo:2021vwj,Isoard:2021peb,Syu:2022cws,Delhom:2023gni,Ciliberto:2024dhl,Gooding:2023xxl}. This is of primary conceptual importance, because entanglement constitutes the core of pair creation in QFTs. Successfully observing entanglement would serve as a crucial benchmark, providing unambiguous confirmation of the quantum origin of the observed excitations.

Observing entanglement, however, presents an extraordinary experimental challenge due to its inherently faint and fragile nature. Indeed, recent claims of observation in analogue spacetimes have been debated \cite{Steinhauer2016,Leonhardt:2016qdi,Steinhauer:2018qzg}, and an independent confirmation of these observations is desired (see \cite{Chen:2021xhd} for observation of quantum correlations in a related scenario, although without an analogue background metric).

Fortunately, technological advancements are progressing at  a fast rate, making the observation of entanglement associated with spontaneous pair creation by dynamical backgrounds attainable in the near future \cite{Jacquet:2020bar}. 
Observing entanglement would rule out any classical explanation of the phenomenon \cite{Martin:2015qta,Green:2020whw,Ashtekar:2020mdv,Agullo:2022ttg}.

The goal of this article is to contribute to this exciting pursuit from the theoretical side. We aim at laying out effective tools for the quantification of the entanglement produced by pair creation, and of the way this entanglement is affected  by various experimental factors, particularly by detrimental effects such as decoherence induced from thermal ambient noise and experimental losses. This quantification is of vital importance for the design of optimal experimental configurations. Although our approach is general and can be applied to  a wide range of experimental setups (see, e.g.,  \cite{Agullo:2021vwj,Brady:2022ffk,Bhardwaj:2023squ,Delhom:2023gni}), in this article we apply it on a current experimental platform which utilizes quantum fluids---Bose-Einstein Condensates (BECs)---to simulate a two-dimensional (2D) expanding universe \cite{Viermann:2022wgw}. Focusing on this  particular experimental platform makes our analysis more concrete and tangible, allowing us to use realistic parameters in our calculations. 

But, most importantly, our focus is motivated by the fact that, as we discuss in this article, the platform presented in \cite{Viermann:2022wgw} is on the verge of observing entanglement originating from cosmological pair creation. Indeed, one of the key outcomes of our analysis is the identification of modest and achievable upgrades to this experiment aimed at reaching optimal configurations for detecting entanglement within the current capabilities and experimental constraints.

The article is organized as follows. In section \ref{sec:BECdynamics} we review the theory behind BEC analogs of expanding universes. In section \ref{sec:Experiment}, we describe the  experimental platform used in \cite{Viermann:2022wgw}, in which classical aspects of acoustic waves compatible with the phenomenon of pair creation were detected after a single expansion ramp. In section \ref{sec:GaussianAndTomography}, we present basic features of the Gaussian state formalism, and explain how to reconstruct the quantum state of the system from time-series measurements of density contrast correlations. We also discuss how to quantify entanglement in this setup. In section \ref{sec:ModelExp}, we discuss our modeling of the proposed experiment, and focus on quantifying the entanglement produced by pair creation. In section \ref{sec:Optimization}, we perform an optimization in experimental parameter space of expansion/contraction experiments aimed at maximizing entanglement detectability, showing how entanglement could be detected with current techniques. Finally, section \ref{sec:Outlook} provides a summary of the main results achieved and includes some discussions on potential future directions.

\section{Quantum sound in analogue expanding universes}\label{sec:BECdynamics}
Expanding universes lead to pair creation for non-conformally invariant quantum fields. Theoretical works have proposed Bose Einstein Condensates (BECs) as analog quantum simulators for real scalar fields in expanding universes \cite{Barcelo2001,Barcelo2003b,Barcelo2003c,Fedichev2004,Fischer2004b,Uhlmann2005,Calzetta2005,Weinfurtner2007,Schuetzhold2009,Prain2010,Bilic2013,Cha:2016esj,GomezLlorente:2019mvt,Bhardwaj:2020ndh,Eckel:2020qee,Tolosa-Simeon:2022umw,Bhardwaj:2023squ}, and exciting experimental realizations of these proposals have already been achieved \cite{eckel_rapidly_2018,Banik:2021xjn,Viermann:2022wgw}.  

In these scenarios, the condensate provides the background spacetime that mimics an expanding universe, and its collective excitations play the role of a quantum field that evolves within. These collective excitations describe quantum sound waves propagating through the condensate, and they display quantum effects that are in analogy with those predicted for real scalar fields in expanding universes.  In this section, we will briefly summarize the theoretical reasoning behind this analogy, emphasizing concept over detail. Readers already familiar with this derivation may want to skip this section.

\subsection{Simulating flat FLRW universes with 2D Bose-Einstein Condensates}

Following \cite{Viermann:2022wgw}, we consider a disk-shaped BEC that is tightly confined in the $z$-direction, with trapping frequency $\omega_z$, making the condensate effectively two-dimensional. The dynamics of this system is described by the Gross-Pitaevskii equation (GPE) \cite{Pitaevskii2016}, which can be derived from the Lagrangian density
\begin{equation}
\begin{split}
     \mathcal{L}=&\frac{\ci\hbar}{2}\lrsq{\Psi^*\partial_t\Psi-(\partial_t \Psi^*)\Psi}-\frac{\hbar^2}{2m}\bs{\nabla}\Psi^*\cdot\bs{\nabla}\Psi\\
     &-V_{\rm ext}\Psi^*{}\Psi{}-\frac{g}{2}\Psi^*{}^2\Psi{}^2\,,
\end{split}
\label{eq:GPELag}
x\end{equation}
where $\Psi$ is a complex scalar field that describes the expectation value of the bosonic field operator, and can be thought of as describing the state of an atom in the condensate. This is the Lagrangian for a  Schr\"odinger field coupled to an external potential $V_{\rm ext}(t,\vx)$ and with a self-interaction term with coupling $g(t)$. Both can explicitly depend on time, and the latter is related to the $s$-wave scattering length of the condensate $\alpha_{\rm s}(t)$ as 
\begin{equation}
    g(t) = \sqrt{\frac{ 8 \pi \omega_z \hbar^{3}}{m}} \alpha_s (t).
    \label{eq:bec.BornApproximation}
\end{equation} 
This quantity can be tuned by exploiting Feshbach resonances using an external magnetic field \cite{Viermann:2022wgw}. The conjugate momentum to $\Psi$ is proportional to $\Psi^*$, meaning  that this theory describes a single real scalar degree of freedom, as it is customary in Schr\"odinger-like theories. This is usually seen by resorting to the hydrodynamical (or Madelung) representation through a field redefinition of the form $\Psi=\sqrt{{n}}\ee^{\ci S}$, in which the complex field $\Psi$ is decomposed in  real density and phase fields, ${n}$ and ${S}$, respectively. According to \eqref{eq:GPELag}, the evolution of density and phase is given by
\begin{equation}
\begin{split}
    &\partial_t n+\frac{\hbar}{m}\vna \lr{n\vna S}=0,\\
    &\hbar\partial_t S+\frac{\hbar^2}{2m}(\vna S)^2+V_{\rm ext}+g n\lr{1-P_{\rm \xi}}=0,
    \end{split}
    \label{eq:StatDensPhase}
\end{equation}
where $P_{\rm \xi}=\xi^2\vna^2\sqrt{n}/\sqrt{n}$ is related to the so called quantum potential, and $\xi\equiv\hbar/\sqrt{2mgn}$ is the healing length of the condensate. It is straightforward to check that ${n}$ and ${S}$ are canonically conjugate variables in the Hamiltonian sense, and thus they  describe a single real scalar field. 

The above equations are analogue to the equations describing an inviscid, irrotational fluid with density $n$ and velocity $\vtv=\hbar\vna S/m$, with a quantum contribution to pressure of the form $g n P_\xi$. This term is proportional to the healing length, and it spoils barotropicity. If the inhomogeneities of the fluid occur over length scales much larger than the healing length, it can be neglected, and the condensate behaves effectively as an inviscid irrotational barotropic fluid. This is called the hydrodynamic (or acoustic) regime of the condensate. When the condensate displays inhomogeneities at scales comparable to the healing length or smaller, the quantum pressure cannot be neglected, the interactions between its microscopic degrees of freedom become relevant, and the hydrodynamic analogy breaks down.

Solutions to the GPE that describe analog Friedmann-Lemaître-Robertson-Walker (FLRW) cosmologies must be isotropic \cite{Tolosa-Simeon:2022umw} and, without loss of generality, one can choose the condensate at rest in the lab frame, so that comoving observers are at rest with respect to the measuring devices. These backgrounds, characterized by $\vna S=0$ and $n(\vx,t)=n(r)$ with $r=|\vx|$, trivially satisfy the continuity equation, while the second equation in \eqref{eq:StatDensPhase} requires
\begin{equation}
    \mu(t)=V_{\rm ext}(r,t)+g(t) n(r)(1-P_\xi(r,t))
\end{equation}
where we have defined the chemical potential $\mu(t)=-\hbar\partial_t S$.

Specific spatial dependence of the  potential $V_{\rm ext}$ lead to spatially flat or curved FLRW analog universes, as shown in \cite{Tolosa-Simeon:2022umw}. In particular, spatially flat geometries correspond to a hard wall potential that vanishes in the region where the condensate is trapped, and lead to a constant density profile. 

\subsection{Sound waves in a BEC and the acoustic metric}

Small perturbations to a fluid consist on density or pressure waves, namely, sound waves. As it is well known \cite{Barcelo:2005fc}, sound waves in inviscid irrotational barotropic fluids are described by a hyperbolic 2nd order PDE with the form of a Klein-Gordon equation in a curved spacetime. The geometry of this spacetime is locally defined by the background fluid. Hence, in the hydrodynamical regime, small perturbations to the condensate are effectively described by a scalar field in a curved spacetime.

To see this, consider a homogeneous solution of \eqref{eq:StatDensPhase} with vanishing velocity and constant density $n_0$, namely $\Psi_0=n_0\ee^{\ci S_0(t)}$. This implies $P_\xi=0$ and requires $\partial V_{\rm ext}/\partial r=0$. Without loss of generality, we can choose $V_{\rm ext}(t)=0$, so that the time-dependence of the chemical potential is dictated by that of the self-coupling as
\begin{equation}
    -\hbar \partial_tS_0=\mu_0(t)=g(t)n_0.
\end{equation}
From here onward, a subscript $0$ indicates that the quantity is associated to the background $\Psi_0$ defined above.

Now, consider small perturbations to this background solution so that 
\begin{equation}
    \Psi(\vx,t)=\Psi_0(t)+\frac{\ee^{\ci S_0(t)}}{\sqrt{2}}\lr{\tilde{\chi}(\vx,t)+\ci \tilde\phi(\vx,t)}\,,
\end{equation}with $\tilde{\chi}$ and $\tilde\phi$ being two real scalar fields parameterizing the perturbations of the complex field $\Psi$. 

As usual, by plugging the above ansatz for $\Psi$ in the Lagrangian \eqref{eq:GPELag}, the dynamics for linear perturbations is described by the quadratic terms in ${\chi}$ and $\phi$. After redefining the fields by $\tilde\phi\to \sqrt{2m}/\hbar\,\phi$ and $\tilde{\chi}\to\chi/\sqrt{2m}$, one arrives at\footnote{We have assumed that the field decays rapidly enough at  spatial infinity, so that  terms proportional to a spatial divergence can be disregarded 
when integrating by parts.}
\begin{equation}
     \mathcal{L}_{2}=-\chi\partial_t\phi-\frac{1}{2}(\vna\phi)^2-\frac{c^2}{2}\lr{\chi^2+\frac{\xi_0^2}{2}(\vna{\chi})^2},
\end{equation}
where we have defined $c(t)=\sqrt{g(t) n_0/m}$. 

As expected, the two scalar fields are not independent degrees of freedom. This can be seen by noting that the conjugate momentum to $\chi$ vanishes and that $\pi_\phi=-\chi$ is the conjugate momentum of $\phi$. Hence, this is a constrained system. 

To solve the constraints, take the Euler-Lagrange equations of the above Lagrangian, which are
\begin{equation}
\begin{split}
    &\chi-\frac{\xi_0^{2}}{2}\vna^2{\chi}=-c^{-2}\partial_t\phi\,,\\
    &\partial_t{\chi}=-\vna^2\phi.
\end{split}
\label{eq:PertEqConst}
\end{equation}
We see that initial data for $\phi$ at a given spatial hypersurface fully specifies $\chi$ and its time derivative. In the hydrodynamical regime, the term proportional to $\xi_0$ can be neglected, so that $\chi$ can be solved in terms of $\partial_t\phi$ and background quantities using the first equation. Plugging this solution in the second equation and defining
\begin{equation}
    \sqrt{-|g|}g^{tt}=c^{-2} 
    \quad\text{and}\quad
    \sqrt{-|g|}g^{ij}=-\delta^{ij} ,
\end{equation}
with $g^{ti}=g^{it}=0$, and where $i,j$ run over Cartesian spatial coordinates $\tx$ and $\ty$, we find the following equation for $\phi$,
\begin{equation}
    \partial_\mu\sqrt{-|g|}g^{\mu\nu}\partial_\nu\phi=0.
    \label{eq:KGEq}
\end{equation}
This is a massless Klein-Gordon (KG) equation for a curved spacetime metric $g_{\mu\nu}$, with determinant $|g|$, and which is defined by the line element
\begin{equation}
    \dd s^2=-\dd t^2+c^{-2}(t)(\dd\tx^2+\dd\ty^2).
    \label{eq:AcMet}
\end{equation}
From the above, we see that $c(t)$ is the speed of propagation of the perturbations through the condensate, or speed of sound, and defines the characteristic curves of the above KG equation, i.e. the soundcone. The metric $g_{\mu\nu}$ is known as the acoustic metric. 

The above line element reproduces a 2+1 flat FLRW cosmology with scale factor 
\begin{equation}
\begin{split}
    a(t)&=\frac{1}{c(t)}=\sqrt{\frac{m}{g(t)n_0}}\\  &=\left(\frac{m^3}{8\pi\omega_z\hbar^3n_0^2}\right)^{1/4}\frac{1}{\sqrt{\alpha_s(t)}}.
\end{split}
\end{equation}
Hence, we can simulate the dynamics of a massless scalar field in a bespoken flat FLRW cosmology by tuning the self-coupling of the atoms in the condensate at will, provided that we stay within the hydrodynamical approximation. Outside this regime, the dispersion relation of phonons is given by the well-known Bogoliubov dispersion relation
\begin{equation}
    \omega_\tk = \pm c\abs{\tk}\sqrt{1+\frac{\xi^2}{2}\tk^2},
\end{equation}
which becomes linear for $\tk\ll\xi^{-1}$. We denote $\tk_{\xi}$ as the wavenumber at which the non-linear correction to the dispersion relation is $10\%$. The hydrodynamical (or acoustic) regime corresponds to $\tk <\tk_{\xi}$.

Let us now discuss some properties of the perturbation equation \eqref{eq:KGEq}. Since it is a KG equation, the symplectic product of two solutions $\phi$ and $\psi$, defined as
\begin{equation}
    \lr{\phi,\psi}=\frac{\ci}{\hbar}\int_{\Sigma_t} \dd^2 \vx \, (\phi^*\pi_\psi-\pi^*_\phi\psi),
    \label{eq:SympProd}
\end{equation}
where $\Sigma_t$ is a constant time 2-dimensional hypersurface, is preserved under time evolution. Here $\pi_\phi$ is the conjugate momentum to $\phi$ which, in the hydrodynamical regime, is given by
\begin{equation}
    \pi_\phi=a^2(t)\partial_t{\phi}.
\end{equation}
The product \eqref{eq:SympProd} defines a (pseudo-)norm in the space of complex solutions to the KG equation. A solution $\phi$ is said to be normalized if it satisfies  $|(\phi,\phi)|=1$. If a solution $\phi$ has positive symplectic norm, its complex conjugate has negative norm, and viceversa.

The homogeneity of the line element \eqref{eq:AcMet} makes it convenient to expand the field in spatial Fourier modes
\begin{equation}
    \phi(t,\vx)=\int \dd^2\vk \, \phi_{\vk}(t,\vx)=\int\frac{\dd^2\vk}{{2\pi}}v_{\tk}(t)\ee^{\ci\vk\cdot\vx},
\end{equation}
where $\tk=|\vk|$. The above will be a solution of the perturbation equation \eqref{eq:KGEq} provided that
\begin{equation}
    \ddot{v}_\tk+2\frac{\dot{a}}{a}\dot{v}_\tk+\frac{\tk^2}{a^2}v_\tk=0,
    \label{eq:ModeEq}
\end{equation}
where dots denote time derivatives. Note that these equations do not distinguish $\vk$ from $-\vk$, which is why the modes $v_\tk$ only depend on the modulus of $\vk$.
Finally, we note that normalization of the modes $\phi_{\vk}$ to a Dirac delta with respect to the symplectic product as $(\phi_{\vk},\phi_{\bs{q}})=\delta^{(2)}(\vk-\bs{q})$ requires
\begin{equation}
    \dot{v}^*_\tk v_{\tk}-v^*_\tk \dot{v}_{\tk}
    =\frac{\ci\hbar}{a^2(t)}.
    \label{eq:NormModes}
\end{equation}
To finish the section, let us analyze the relation between our field perturbations and observables. Perturbations can also be written in terms of density and phase variables, which describe observable quantities. Following the definition of $\chi$ and $\phi$, if we perturb the condensate in density-phase variables as $n=n_0+n_1$ and $S=S_0+S_1$, these perturbations are related to our fields as
\begin{equation}
    \chi=\sqrt{\frac{m}{n_0}}n_1
    \qquad\text{and}\qquad
    \phi=\sqrt{\frac{\hbar n_0}{m}}S_1
\end{equation}
Hence, we can write the density contrast to linear order in perturbations as
\begin{equation}
    \delta_c(t,\vx)\coloneqq\frac{n(t,\vx)-n_0}{n_0}=-\frac{\pi_\phi}{\sqrt{m n_0}}=-\frac{a^2}{\sqrt{m n_0}}\partial_t\phi.
    \label{eq:DensCont}
\end{equation}
where the last equation only holds in the hydrodynamical regime, so that the $\xi_0$ term in \eqref{eq:PertEqConst} can be neglected. Note that measurements of $\delta_c$ and its time evolution allow to reconstruct the field $\phi$ at all times through Hamilton's equations, which allow to write  $\phi$ in terms of $\partial_t \pi_\phi$. This will be useful in section \ref{sec:Tomography}.

\subsection{Quantization, time-evolution, and pair creation}\label{sec:QuantizationTimeEvol}

Because condensates are formed at low temperatures, quantum effects are relevant to their description. Thus, experiments can be sensitive to the quantum nature of the perturbations described above. In the following, we summarize the  quantization of the perturbations, paralleling  the familiar  quantization  of a scalar field in an expanding universe.

Quantization proceeds as usual, by promoting the scalar field and its conjugate momentum to operators $\hat\Phi(t,\vx)$ and $\hat\Pi(t,\vx)$ which satisfy canonical commutation relations $$\lrsq{\hat\Phi(t,\vx),\hat\Pi(t,\vy)}=\ci\hbar\delta^{(2)}(\vx-\vy).$$ 

 The perturbation equation is a linear 2nd order PDE with real coefficients, meaning that, if $\phi_{\vk}(\vx,t)$ is a solution, so is $\phi^*_{\vk}(\vx,t)$. For $\phi_{\vk}(\vx,t)$ of the form $\phi_{\vk}(\vx,t)=v_\tk(t)\, \ee^{\ci\vk\cdot\vx}$ with $v_\tk(t)$ satisfying \eqref{eq:NormModes}, the set of pairs 
 $\{\phi_{\vk},\phi_{\vk}^*\}$, for all $\vk$ forms an orthonormal basis of the space of complex solutions to \eqref{eq:KGEq}. The construction of such a basis requires a choice of the functions $v_\tk(t)$ satisfying \eqref{eq:ModeEq} and the normalization  \eqref{eq:NormModes}. In non-expanding geometries, time translational invariance makes a preferred choice available, namely positive frequency exponentials  $v_\tk(t)\propto \ee^{-\ci\omega t}$. In expanding geometries there is no global preferred choice of basis. However, when the scale factor becomes constant in the past and future, we can use the asymptotic time translations symmetries to construct the IN and OUT basis, as we describe below. 

Once a basis has been chosen, one can write a quantum representation for the field and momentum operators in the standard way. Namely, one defines a pair of non-Hermitian operators for each $\vk$, by ``projecting'' the field operator onto the basis elements
\begin{equation}
\begin{split}
     &\bk:=(\phi_{\vk},\hat{\Phi})=\frac{\ci}{\hbar}\int_{\Sigma_t} \dd^2 \vx \,( \phi_{\vk}^*\hat\Pi-\pi^*_{\vk}\hat\Phi),\\
     &\bmkd:=-(\phi^*_{\vk},\hat{\Phi})=-\frac{\ci}{\hbar}\int_{\Sigma_t}\dd^2 \vx \,  (\phi_{\vk}\hat\Pi-\pi_{\vk}\hat\Phi),
\end{split}
    \label{eq:DefCreationOp}
\end{equation}
where $\pi_{\vk}=\dot a^2(t)\, \dot \phi_{\vk}$. These operators satisfy
\begin{equation}
\lrsq{\hb_{\vk},\hb_{\bs{q}}^\dagger}=\delta^{(2)}(\vk-\bs{q})\,.   
\end{equation}
A representation of the field operator is then given by
\begin{equation}
\begin{split}
     \hat\Phi(t,\vx)=\int \frac{\dd^2\vk}{2\pi}\ee^{\ci\vk\cdot \vx}\, \hat\Phi_{\vk}(t),
     \label{eq:FourierExpQuantumField}
\end{split}
\end{equation}
where $\hat\Phi_{\vk}(t)=v_{\tk}(t)\bk+v_{\tk}^*\bmkd$. This is a quantum system with infinitely many degrees of freedom, each described by pairs of canonical variables $\hat\Phi_{\vk}(t)$ and $\hat\Pi_{\vk}(t)$. In the hydrodynamical regime, $\hat\Pi_{\vk}$ is given by
\begin{equation}
\hat{\Pi}_{\vk}=a^2(t)\dot{\hat{\Phi}}_{\vk}\,.
\label{eq:QuantumFieldMom}
\end{equation}
Alternatively, we can also describe our system using pairs of non-Hermitian operators $\bk$ and $\bkd$.
The Fock space of the theory is constructed by acting with creation operators $\bk^{\dagger}$ on the Fock vacuum, defined as the state annihilated by $\bk$ for all $\vk$. 

Let us restrict to scale factors which become static in the past and future, equal to  $\ai$ for $t<t_{\rm i}$ and to $\af$ for $t>t_{\rm f}$, respectively. 

The IN representation is defined by choosing $v_\tk(t)$ satisfying
\begin{equation}
\begin{split}
    &v_{\tk}(t_0<t_{\rm i})=\sqrt{\frac{\hbar}{2\omi}}\frac{\ee^{-\ci\omi t_0}}{\ai}\,,
    \\
    &\dot{v}_{\tk}(t_0<t_{\rm i})=-\ci\omi\sqrt{\frac{\hbar}{2\omi}}\frac{\ee^{-\ci\omi t_0}}{\ai}\,,
\end{split}
\label{eq:EarlyTimeInCond}
\end{equation}
where $\omi=c_{\rm i}\tk=\tk/\ai$, and where the amplitude is chosen to satisfy \eqref{eq:NormModes}. The Fock vacuum defined by these modes is called the IN vacuum state $\vacin$. It corresponds to the ground state of the system before the expansion begins. 

The OUT representation is similarly defined. We will use the symbol $u_\tk$ instead of $v_\tk$ to denote the OUT modes. They are the unique solutions to  \eqref{eq:ModeEq} satisfying
\begin{equation}
\begin{split}
    &u_{\tk}(t_0>t_{\rm f})=\sqrt{\frac{\hbar}{2\omf}}\frac{\ee^{-\ci\omf t_0}}{\af}\,,
    \\
    &\dot{u}_{\tk}(t_0>t_{\rm f})=-\ci\omf\sqrt{\frac{\hbar}{2\omf}}\frac{\ee^{-\ci\omf t_0}}{\af}\,,
\end{split}
\label{eq:LateTimeInCond}
\end{equation}
where $\omf=c_{\rm f}\tk=\tk/\af$. These modes define a set of OUT creation and annihilation operators $\{\bkout,\bkdout\}$ and an OUT vacuum $\vacout$, which will in general differ from $\vacin$. To quantify this difference, we first write the OUT mode basis as a linear combination of IN modes
\begin{equation}
\begin{split}
    &u_\tk=\alpha_\tk v_\tk+\beta_\tk v^*_\tk,\\
    &u^*_\tk=\alpha_\tk^* v^*_\tk+\beta_\tk^* v_\tk.
\end{split}
\label{eq:BogoModes}
\end{equation}
Multiplying the first equation by $\ee^{\ci\vk\cdot\vx}$ and the second equation by its conjugate, and taking  symplectic products suitably, one can show two things. First of all, using the definition of creation and annihilation operators \eqref{eq:DefCreationOp},  the relation between IN and OUT operators reads
\begin{equation}
\begin{split}
    &\bkout=\alpha_\tk^* \bkin-\beta_\tk^* \bmkdin,\\
    &\bkdout=\alpha_\tk \bkdin-\beta_\tk\bmkin.
\end{split}\label{bogt}
\end{equation}
Secondly, the coefficients of the transformation must satisfy
\begin{equation}
    \begin{split}
        &|\alpha_\tk|^2-|\beta_\tk|^2=1,
        \\
        &\alpha_\tk^*\beta_{\tk}-\beta_\tk^*\alpha_{\tk}=0.
    \end{split}
    \label{eq:BogoConst}
\end{equation} 
Hence, this is a Bogoliubov transformation. Note that the IN and OUT vacua will correspond to different physical states when $\beta_\tk\neq0$. Indeed, if we prepare the system in the state  $\vacin$ at $t<t_{\rm i}$,   OUT observers in the asymptotic future will observe  a particle content
\begin{equation}
    \expval{\hat{N}^{\rm out}_{\vk}}_{\rm in}={}_{\rm in}\bra{0}\bkout\bkdout\vacin=|\beta_{\tk}|^2.
\end{equation}
Equations \eqref{bogt}  inform us that evolution only mixes $\vk$ and $-\vk$ modes, as expected in a homogeneous and isotropic universe. The evolution of a pair $(\vk,-\vk)$ of modes decouples from the evolution of any other pair of modes.  Hence, we can focus on one pair  $(\vk,-\vk)$ at a time. Let 
\begin{equation}
    \hat{\bs{B}}_{\vk}=\lr{\bk,\bkd,\bmk,\bmkd}^\top
    \label{eq:PhaseSpaceVec}
\end{equation}
be the annihilation and creation operators for one such pair. With this notation, we encode the evolution in the following equation
\begin{equation}
    \hat{\bs{B}}_{\vk}^{\rm out}=\bS_{\tk} \cdot\hat{\bs{B}}_{\vk}^{\rm in}\,,
    \label{eq:QuantEvol}
\end{equation}
where
\begin{equation}
    \bS_\tk=
    \begin{pmatrix}
        \alpha_\tk^* & 0 & 0 & -\beta_\tk^*\\
        0 &  \alpha_\tk & - \beta_\tk & 0\\
        0 & -\beta_\tk^* & \alpha_\tk^* & 0\\
        -\beta_\tk & 0 & 0 & \alpha_\tk\\
    \end{pmatrix}
    \label{eq:SmatrixTh}
\end{equation}
is  a symplectic matrix\footnote{In the sense that $\bS\cdot\bs{\Omega}\cdot\bS^\top=\bs{\Omega}$, with $\bs{\Omega}=\begin{pmatrix}
    0 & 1\\
    -1 & 0
\end{pmatrix}
\oplus
\begin{pmatrix}
    0 & 1\\
    -1 & 0
\end{pmatrix}$.} 
providing the relation between in and out operators. 

In order to find the elements of $\bS_\tk$, one can proceed in several manners. One of them is to first find a numerical solution to the mode equation with initial conditions at late times $t_0>t_{\rm f}$ given by \eqref{eq:LateTimeInCond}. Then evolve it back in time and write it as a linear combinations of IN plane waves of the form \eqref{eq:EarlyTimeInCond}. The coefficients of the combination are the Bogoliubov coefficients as defined in \eqref{eq:BogoModes}. We will later use this procedure to compute  observables of interest, including entanglement, for particular examples of dynamical universes that are relevant for current and future experiments. 

\section{Experimental realization of an expanding universe in a BEC}
\label{sec:Experiment}

In this section we will briefly describe a recent experiment \cite{Viermann:2022wgw} in which the condensate is tuned to mimic a universe where the scale factor interpolates between an initial value $\ai$ and a larger final value $\af$ with a polynomial expansion rate $a(t)\sim t^\gamma$. The expansion lasts a period of time  $\Delta t=t_{\rm f}-t_{\rm i}$. In section \ref{sec:ProdEnt}, we will propose feasible modifications to this experiment targeting optimization in the detectability of the entanglement produced by pair creation.

The experiment is carried out with a condensate made of ${}^{39}$K atoms which is tightly confined into a 2-dimensional plane and then radially confined in a region where the condensate is approximately homogeneous. The scattering length of the condensate is then controlled through a magnetic field in the vicinity of a Feshbach resonance of ${}^{39}$K, and ramped from an initial value of $\alpha_{s, \text{i}}=400 \, a_{\rm B}$ to a value of $\alpha_{s, \text{f}}=50 \, a_{\rm B}$, where $a_{\rm B}$ is the Bohr radius. This amounts to an increase of the scale factor by $\af=\sqrt{8}\ai$. The expansion curve is given by a scattering length of the form\footnote{This translates into a scale factor of the form $a(t)=a_{\rm i}\lr{\frac{t+t_0}{t_0}}^{\gamma}$}
\begin{equation}
    \alpha_{s}(t)=\alpha_{s,\rm i} \lr{\frac{t+t_0}{t_0}}^{-2\gamma}
\end{equation}
with $t_0=\Delta t \lr{(\alpha_{s,\rm i}/\alpha_{s, \rm f})^{1/2\gamma}-1}^{-1}$. The expansion interval $\Delta t=t_{\rm f}-t_{\rm i}$ is chosen to be \SI{1.5}{\milli\second} and \SI{3}{\milli\second}, with $\gamma$ values of $0.5$, 1, and 1.5, so the expansion is decelerating, constant or accelerating, respectively. An external potential is used so that the shape of the condensate does not change during the experiment.

After the expansion, the condensate is allowed to evolve freely, and measurements of the condensate density are performed at different times after the expansion, with a resolution in time of $\approx\SI{0.25}{\milli\second}$. The background density $n_0$ is obtained by averaging the density at each pixel $(\tx,\ty)$ through all repetitions of the experiments with the same preparation . Then, the density contrast at each pixel $\vx$ at time $t\geq t_{\rm f}$ after the expansion is computed for each realization as $\delta_c(t,\vx)=n(t,\vx)/n_0-1$,  where $n(t,\vx)$ is the measured value of the density in that pixel at time $t$. Then, two-point density contrast correlations are computed at different times and averaged over all the realizations with identical initial preparation. Due to homogeneity and isotropy, the resulting correlation function depends only on the distance between points, not their precise location. Hence, the obtained correlations are averaged over all pairs of points at the same distance $L$. The resulting correlation functions depend on $L$ and the time after the expansion $t$, and they encode information about the expansion history.

The expansion produces sound waves that propagate through the condensate. These leave a characteristic imprint in the correlation functions, in the form of an anti-correlation dip followed by a small correlation peak after the dip. The soundspeed of the condensate after the expansion can be measured either from the evolution of correlations as half the speed of propagation of the correlation features, or from wavepacket propagation. Due to homogeneity and isotropy, phonons are produced only in pairs which travel in opposite directions at the speed of sound in the hydrodynamical regime $\tk\lesssim \tk_{\xi}$, so that the distance between them increases as twice the speed of sound. This method leads to a measured speed of sound value of $2c=$\SI{2.5(1)}{\micro\meter \,\milli\second^{-1}}, in agreement with the value computed from the scattering length after the expansion. Observed modes lie within the hydrodynamical regime with a healing length of $\xi_0=$\SI{0.5}{\micro\meter}.

The resulting correlation function is then transformed into the spectrum of fluctuations $S_\tk$ through a Hankel transform of order $0$ (see section \ref{sec:Tomography}). The spectrum is then fitted to the form it should have for a freely evolving condensate $S_\tk=B_\tk+A_\tk\cos\big(c \tk (t-t_{\rm f})+\theta_\tk\big)$, with offset $B_\tk$, amplitude $A_\tk$, and phase $\theta_\tk$ as fitting parameters. The obtained values for $A_\tk$ and $\theta_\tk$ are then compared with theoretical predictions, finding a good agreement with data assuming 1) an initial  thermal state at $T=\SI{40}{\nano\kelvin}$, 2) a value for the speed of sound after the expansion of $c=$\SI{1.2}{\micro\meter \,\milli\second^{-1}} at a final scattering length of $50 \, a_{\rm B}$, 3) an initial scattering length of $350 \, a_{\rm B}$ and 4) dynamics given by \eqref{eq:KGEq}.  The temperature is determined experimentally to be $T=$\SI{60(10)}{\nano\kelvin} through a fit to the observed initial density profile. Though the value of $A_\tk$ is sensitive to temperature (and more generally, to the details of the initial state), the value of the phase $\theta_\tk$ is not. Hence, agreement in $\theta_\tk$ is taken as a good indicator that the observed data corresponds to the process of pair creation due to the expansion history implemented experimentally in \cite{Viermann:2022wgw}.

In the remainder of this article, we present a set of theoretical tools to quantify the entanglement associated with the pair-creation process. Building upon the findings of \cite{Viermann:2022wgw}, we identify modifications to the experimental setup and regions in the parameter space that optimize entanglement production, and quantify its detectability.

\section{Gaussian states and quantum tomography from observations}\label{sec:GaussianAndTomography}

This section describes how to perform quantum state tomography from time-series measurements of the density contrast after the expansion/contraction phase has taken place. To that end, we will first briefly introduce Gaussian states along with some tools that make this task easier. These are the same tools we will use  in subsequent sections to quantify  entanglement.
\subsection{Gaussian states}
General quantum states are completely characterized by the value of all $n$-point functions of their canonical variables. In our case, these correspond to quantum averages of all the possible products of components of the vector $\hat{\bs{B}}$ defined in \eqref{eq:PhaseSpaceVec}, in all possible orderings. However, in many experiments the state of the system is well approximated by a Gaussian state. This family of states poses a remarkably simple structure of $n$-point correlation functions. 

Gaussian states form a special class of quantum states characterized by the fact that they are entirely determined by their one- and two-point correlation functions. This information can be encoded in the mean $\bs{\mu}$ and covariance matrix $\bs{\sigma}$ of the state. In our case, these are defined from the vector of canonical operators $\hat{\bs{B}}$ as
\begin{equation}
\begin{split}
    &{\mu}^i=\expval{\hat{{B}}^i},\\
    &{\sigma}^{ij}=\expval{\{\hat{B}^i-\mu^i,\hat{B}^j-\mu^j\}},
\end{split}
\label{eq:MeanAndCov}
\end{equation}
where $i$ and $j$ run over components of $\hat{\bs{B}}$, and $\{\cdot,\cdot\}$ stands for anticommutator---the antisymmetric part of the two-point function contains no information on the state, since it just encodes the canonical commutation relations. 

Note that the components of $\hat{\bs{B}}$ can be expressed in any basis of canonical operators. For instance, we could alternatively use $\hat{\bs{B}}_{\vk}^{(\Phi)}=(\hat{{\Phi}}_{\vk},\hat{{\Pi}}_{\vk},\hat{{\Phi}}_{-\vk},\hat{{\Pi}}_{-\vk})^\top$. In our system, the change of canonical basis from creation/annihilation to field/momentum operator basis is given by the matrix
\begin{equation}
    \bs{P}_\tk=
    \begin{pmatrix}
   v_\tk & 0 & 0 & v_\tk^* \\
  a^2 \dot{v}_\tk & 0 & 0 & a^2\dot{v}_\tk^* \\
    0 & v_\tk^* & v_\tk & 0 \\
    0 &a^2 \dot{v}_\tk^* &a^2\dot{v}_\tk & 0
    \end{pmatrix},
\end{equation}
so $\hat{\bs{B}}_{\vk}^{(\Phi)}=\bs{P}_\tk \cdot\hat{\bs{B}}_{\vk}$, and therefore $\bs{\sigma}^{(\Phi)}_{\vk}=\bs{P}_\tk \cdot\bs{\sigma}_{\vk}\cdot\bs{P}_\tk^\top$, where $\bs{\sigma}^{(\Phi)}_{\vk}$ and $\bs{\sigma}_{\vk}$ denote the covariance matrix in the operator bases $\hat{\bs{B}}_{\vk}^{(\Phi)}$ and 
$\hat{\bs{B}}_{\vk}$, respectively.

For Gaussian states, all higher-order $n$-point functions are expressible in terms of the mean and covariance matrix, greatly simplifying computations.  Gaussian states include vacuum, thermal, coherent, or squeezed states; which typically describe well the outcomes of the experiments we are considering.

The Gaussian character of a state is preserved by linear evolution (i.e., evolution generated by a Hamiltonian that is quadratic in the canonical variables). Indeed, from \eqref{eq:QuantEvol}, a quantum system initially prepared in a Gaussian state at $t_0$, characterized by $\bs{\mu}(t_0)$ and $\bs{\sigma}(t_0)$,  evolves to another Gaussian state with
\begin{equation}
\begin{split}
    &\bs{\mu}(t)=\bS(t-t_0)\cdot\bs{\mu}(t_0),\\
    &\bs{\sigma}(t)=\bS(t-t_0)\cdot\bs{\sigma}(t_0)\cdot\bS^\top(t-t_0)\,,
\end{split}
\end{equation}
where $\bS(t-t_0)$ is a symplectic matrix. 

Another relevant property of Gaussian states is that, if a bipartite system $AB$ is in a Gaussian state $(\bs{\mu}_{AB},\bs{\sigma}_{AB})$, then the reduced states of each subsystem are also Gaussian states, denoted as $(\bs{\mu}_{A},\bs{\sigma}_{A})$ and $(\bs{\mu}_{B},\bs{\sigma}_{B})$. The mean vector $\bs{\mu}_{A}$ consists of the elements in $\bs{\mu}_{AB}$ that correspond to the modes of subsystem $A$, while the covariance matrix $\bs{\sigma}_{A}$ is defined by the elements of $\bs{\sigma}_{AB}$ describing correlations solely between the modes of subsystem $A$ (with an analogous definition for subsystem~$B$).

\subsection{State tomography from time-series measurements of density contrast}\label{sec:Tomography}

This section describes how to reconstruct the quantum state of acoustic perturbations from experimental data, along with the assumptions behind such reconstruction.

In the experiments under consideration, the relevant information is gathered through measurements of the density at different times. The quantum sound waves correspond to density fluctuations over the mean value $n_0$. Repeated experiments with identical values for the parameters permit the measurement of both the mean density and the density contrast $\hat{\delta}_c(t,\mathbf{x})$ (defined in \eqref{eq:DensCont}). Averages over a large number of identical experiments also allow for the computation of $n$-point correlation functions of $\hat{\delta}_c(t,\mathbf{x})$. Furthermore, time-series measurements permit the measurement of correlation functions at different times, from which time derivatives can be obtained using a finite-differences method.

If we assume that the initial state of acoustic perturbations is Gaussian and their dynamics are described by the linear equations of motion discussed in Section \ref{sec:BECdynamics}, the final state is also Gaussian. Gaussianity implies that the one- and two-point correlation functions of the field and momentum exhaust the information in the quantum state. The one-point function vanishes by definition of density contrast $\hat{\delta}_c$, hence we focus attention in the two-point function. Knowledge of the equations of motion allows us to relate the time derivative of the field to its conjugate momentum. Altogether, measurements $\langle \hat{\delta}_c(t,\mathbf{x}) \hat{\delta}_c(t,\mathbf{x}') \rangle$ suffice to reconstruct the quantum state.

The assumptions of Gaussianity and the validity of the equations of motion are already needed to interpret this experiment, so no additional assumptions are introduced in the state reconstruction. Furthermore, one could test Gaussianity and the linearity of the equations of motion by measuring higher-order correlation functions, such as the three- and four-point functions.

The state reconstruction is more conveniently done using Fourier space. In the following, we spell out some intermediate steps needed to arrive at our final expression \eqref{eq:CovarianceMatrixOffAmpPhase} for the covariance matrix of the final state in terms of quantities that can be directly extracted from the experiments, and which form the basis of the experimental analysis done in \cite{Viermann:2022wgw}.

Using the relation between the conjugate momentum $\hat{\pi}_{\phi}$ and the density contrast \eqref{eq:DensCont}, and employing the field equations in Hamiltonian form, we find
\begin{equation}
    \begin{split}
        &\expval{\{\hat{\Pi}_{\vk}(t)\hat{\Pi}_{\vq}(t'))\}}=m n_0 \expval{\{\hat{\delta}_c{}_{\vk}(t)\hat{\delta}_c{}_{\vq}(t')\}}\,,
        \\
         &\expval{\{\hat{\Phi}_{\vk}(t),\hat{\Pi}_{\vq}(t')\}}=-\frac{m n_0}{{\tk^2}}\frac{\dd }{\dd t}\expval{\{{\hat{\delta}}_c{}_{\vk}(t)			\hat{\delta}_c{}_{\vq}(t')\}}\,,\\
        &\expval{\{\hat{\Phi}_{\vk}(t),\hat{\Phi}_{\vq}(t')\}}=\frac{m n_0}{\tk^4}\frac{\dd }{\dd t}\frac{\dd}{\dd t'}\expval{\{\hat{\delta}_c{}_{\vk}(t)\hat{\delta}_c{}_{\vq}(t')\}}.
    \end{split}
    \label{eq:DensFieldCorrel}
\end{equation}

Taking the limit $t'\to t$, these relations provide the elements of the covariance matrix of the system at a given instant.\footnote{It is important to note that, although the above relations only hold in the hydrodynamical regime, where the dynamics is described by \eqref{eq:KGEq}, a similar set of relations can be obtained for any other set of dynamical laws governing the evolution of the system. Hence, the reconstruction of the quantum state of the system from time-series measurements of the density contrast is rather general and not restricted to the hydrodynamical regime.}

Following \cite{Tolosa-Simeon:2022umw}, if the system is assumed to be in a homogeneous and isotropic state, the density contrast two-point correlation function only depends on $L = |\vx - \vy|$, and it can be conveniently encoded in the {\em spectrum} $S_\tk$, defined as the Hankel transform of weight $0$ of the density-contrast correlations:
\begin{equation}
\begin{split}
S_{\tk}(t)=&\frac{\pi m n_0}{{\hbar} \tk a(t)}\int_{\mathbb{R}^+} \dd L\,  L \, J_{0}(\tk L)\expval{\left\{\hat{\delta}_c(t,\vx),\hat{\delta}_c(t,\vy)\right\}},
\label{eq:DefinitionSpectrum}
\end{split}
\end{equation}
Using orthonormality properties of Bessel functions, we obtain
\begin{equation}
    \begin{split}
        &\expval{\{\hat{\Pi}_{\vk},\hat{\Pi}_{\vq}\}}
        =
        2{\hbar}\tk a(t)S_\tk\, {\delta^{(2)}(\vk+\vq)}\, ,
        \\
       & \expval{\{\hat{\Phi}_{\vk},\hat{\Pi}_{\vq}\}}
        =
        -\frac{{\hbar}a(t)}{\tk}\dot{S}_\tk \, {\delta^{(2)}(\vk+\vq)}\, ,
	\\
        &\expval{\{\hat{\Phi}_{\vk},\hat{\Phi}_{\vq}\}}
        =\lr{
        \frac{{\hbar}a(t)}{\tk^2}\ddot{S}_\tk+\frac{2{\hbar}}{\tk a(t)}\dot{S}_\tk} \, {\delta^{(2)}(\vk+\vq)}.
    \end{split}
    \label{eq:CovMatElemExp}
\end{equation}
The presence of $\delta^{(2)}(\vk + \vq)$ reminds us that the dynamics only mixes modes with wavenumber $\vk$ and $-\vk$, permitting us to focus on one pair $(\vk, -\vk)$ at a time.

From this, the covariance matrix of  a pair $(\vk,-\vk)$ of density perturbations reads, in the field/momentum operator basis:
\begin{equation}
    \bs{\sigma}_{\vk}^{(\Phi)}
    =
    \begin{pmatrix}
    0 & 0 &  \frac{{\hbar}a}{\tk^2}\ddot{S}_\tk+\frac{2{\hbar}}{\tk a}\dot{S}_\tk &  -\frac{{\hbar}a}{\tk}\dot{S}_\tk\\
    0 & 0 & -\frac{{\hbar}a}{\tk}\dot{S}_\tk & 2{\hbar}\tk a S_\tk \\
     \frac{{\hbar}a}{\tk^2}\ddot{S}_\tk+\frac{2{\hbar}}{\tk a}\dot{S}_\tk  &  -\frac{{\hbar}a}{\tk}\dot{S}_\tk\ & 0 & 0 \\
     -\frac{{\hbar}a}{\tk}\dot{S}_\tk\ & 2{\hbar}\tk a S_\tk & 0 & 0
    \end{pmatrix}.
\label{eq:CovarianceMatrixSk}
\end{equation}
Finally, we note that, at any time $t\geq t_{\rm f}$ after the expansion, when the system is undergoing free evolution,  the spectrum can be conveniently parameterized as \cite{Tolosa-Simeon:2022umw}
\begin{equation}
    S_\tk=B_\tk+A_\tk \cos\lrsq{2\omega^{\rm f}_\tk (t-t_{\rm f})+\theta_\tk},
\label{eq:SpectrumFit}
\end{equation}
where $A_\tk\geq0$ and $B_\tk\geq1/2$ (with the equalities holding true if the system were in the OUT vacuum), and  $A_\tk\leq B_\tk$. 

Recall from section \ref{sec:Experiment} that the quantities $A_\tk$, $B_\tk$ and $\theta_\tk$, known as amplitude, offset, and phase, are directly extractable from experimental data, and form the basis of the analysis done in \cite{Viermann:2022wgw}. From them, the covariance matrix of density perturbations at any time $t \geq t_{\rm f}$ is, when written in the creation/annihilation operator basis:
\begin{equation}
   \bs{\sigma}_{\vk}^{\text{out}} = 
   2
    \begin{pmatrix}
    0 & B_\tk & -A_\tk\ee^{-\ci\theta_k} & 0\\
    B_\tk &  0 & 0 & -A_\tk\ee^{\ci\theta_k} \\
   -A_\tk\ee^{-\ci\theta_k} & 0 & 0 & B_\tk \\
    0 & -A_\tk\ee^{\ci\theta_k} & B_\tk & 0
    \end{pmatrix}.
\label{eq:CovarianceMatrixOffAmpPhase}
\end{equation}
From this covariance matrix, one can reconstruct the reduced density matrix for each
pair $(\vk,-\vk)$, following the tools outlined in \cite{serafini17QCV}. However, such a step is not necessary, since all physical predictions can be easily extracted directly from the covariance matrix $\bs{\sigma}_{\vk}$. 

To summarize, by assuming Gaussianity, the form of the equations of motion, and homogeneity, the experimental data collected in \cite{Viermann:2022wgw} suffices to perform full state tomography. Furthermore, these assumptions can be tested with the available data. The information encoded in the quantum state is compactly represented by the covariance matrix \eqref{eq:CovarianceMatrixOffAmpPhase} for each pair of Fourier modes $(\vk,-\vk)$. The  reconstruction of the final state does not require any additional assumptions or extra knowledge about the experiment, such as the temperature of the initial state, experimental losses, or detector efficiencies.

Importantly, as discussed in more detail in section \ref{entanglement}, knowledge of $\bs{\sigma}_{\vk}^{\text{out}}$ allows for the identification and quantification of the entanglement in the final state. This approach readily accounts for all decoherence effects arising from losses and ambient thermal noise that may be present in the experiment.


\section{Theoretical modeling of the experiments}
\label{sec:ModelExp}
If entanglement is present in the final state, a comparison against theoretical predictions is needed to identify its physical origin, particularly whether it stems from pair production.

This section is devoted to outlining a theoretical model that describes experiments involving dynamical universe analogs, using the tools discussed in previous sections. From the model, we will derive theoretical predictions that can be quantitatively contrasted with experimental data. Special attention will be given to quantifying the expected entanglement resulting from the pair creation process, and how it changes as a function of $\tk$.

We will consider the particular case of a 2D condensate with $N \approx 45 000$ atoms confined in a region of radius $25 \, \mu\text{m}$, with trapping frequency $\omega_z=2\pi \times 1.6 \, \text{kHz}$ and a constant density $n_0 \simeq 5.73 \times 10^8 \, \text{cm}^{-2} $, in line with current realizations \cite{NewPaper}.

\subsection{Modeling different expansion histories}
\label{sec:RampModels}

\begin{figure*}[t!]
    \centering
    \includegraphics[width=0.9\textwidth]{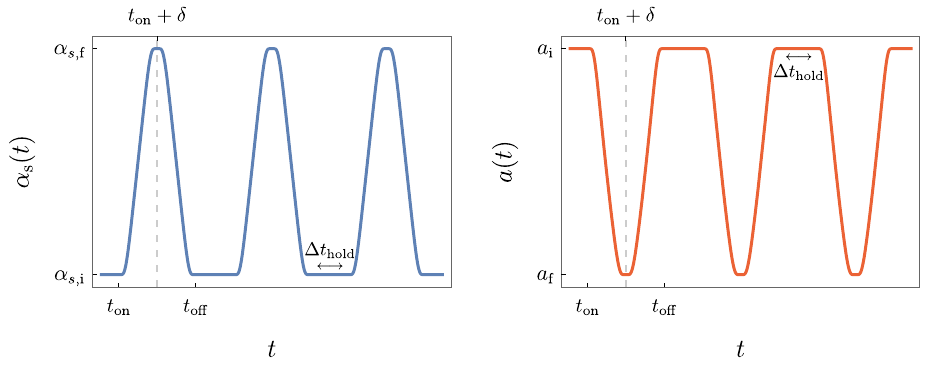}
    \caption{
    Left panel: Scattering length $\alpha_s(t)$ as a function of time, made up of a sequence of cusps with duration $2\delta$, with $\delta = 0.5\, \text{ms}$. The cusps are separated by an interval $\Delta t_{\text{hold}}$, during which the scattering length is kept constant and equal to $\alpha_{s,\text{i}}$. We have highlighted the limits of the static regimes before and after the first cusp, using the labels $t_{\text{on}}$ and~$t_{\text{off}}$. Right panel: Corresponding scale factor $a(t)$. Note the inverse relationship, $a(t) \propto 1/\sqrt{\alpha_s(t)}$. Although it may not be immediately apparent due to the value of the parameters chosen in these plots, both curves are infinitely differentiable.}  \label{fig:ScatteringLength}
\end{figure*}

We will considered different expansion/contraction histories, all of which  can be achieved  with the techniques presented in \cite{Viermann:2022wgw} (see \cite{NewPaper}). 

In section \ref{sec:Experiment}, we discussed power-law scale factors of the form $a(t) \sim t^{\gamma}$, which drive the expansion from $t_{\rm i}$ to $t_{\rm f}$. However, we will see later that the entanglement produced by such power-law scale factors is too small to be detected with current experimental capabilities or modest improvements thereof (see Figure \ref{fig:LinearRampLN} in section \ref{sec:ProdEnt}). Instead, we propose using a scale factor with repeated expansion and contraction cycles. This is advantageous, because resonant effects significantly enhance particle and entanglement production. Such scenarios have been typically modeled by sudden changes in the scattering length \cite{Schmidt:2024zpg}. Our analysis will consider smooth changes to avoid spurious effects due to instantaneous transitions.

More precisely, we model the expansion history using a scattering length composed of a sequence of $n$ ``cusps" joined by periods of constant $\alpha_s = \alpha_{s,\text{i}}$ (see the left panel in Figure~\ref{fig:ScatteringLength}; the right panel shows the corresponding scale factor). Each cusp is modeled by the function:
 
\begin{eqnarray}
\alpha^{\rm cusp}_s(t) = \alpha_{s,\text{i}} &+& \left(\alpha_{s,\text{f}} - \alpha_{s,\text{i}}\right)\big[\Theta_{\delta}(t - t_{\text{on}} - \delta/2) \nonumber \\
    &- & \Theta_{\delta}(t - t_{\text{off}} + \delta/2)\big],
\label{eq:OneCuspScatteringLength}
\end{eqnarray}
where $\Theta_{\delta}(t) = \Big[1 + \tanh\big(\coth[\pi(\frac{1}{2} - \frac{t}{\delta})]\big)\Big]/2$. The duration of each cusp is $2\delta$ (i.e., in Figure~\ref{fig:ScatteringLength}, $t_{\text{off}} = t_{\text{on}} + 2\delta$). Hence, $\delta$ controls the ``steepness" of each cusp, with the steepness increasing as $\delta$ approaches 0.

The time interval between consecutive cusps, denoted as $\Delta t_{\rm hold}$, during which $\alpha_s(t)$ is kept constant and equal to $\alpha_{s,\text{i}}$, is a free parameter.

The expansion/contraction history we consider is determined by the parameters $\alpha_{s,\text{f}}$, $\frac{\alpha_{s,\text{f}}}{\alpha_{s,\text{i}}}$, $\delta$, $\Delta t_{\rm hold}$, and $n$. The value of $\alpha_{s,\text{f}}$ dictates $a_{\rm f}$, the minimum value of the scale factor, while $\frac{\alpha_{s,\text{f}}}{\alpha_{s,\text{i}}}$ determines the amount of expansion/contraction during each cusp. Particle production is only sensitive to the ratio $\frac{\alpha_{s,\text{f}}}{\alpha_{s,\text{i}}}$, not to the absolute value of $\alpha_{s,\text{f}}$, although the latter is experimentally relevant. We will fix it to $\alpha_{s,\text{f}} = 400\aB$ for our analysis. 

In the following, we will study particle and entanglement production for different values of the parameters $\frac{\alpha_{s,\text{f}}}{\alpha_{s,\text{i}}}$, $\delta$, $\Delta t_{\rm hold}$, and $n$. Experimentally, values of $\delta$ as small as 0.1 ms are achievable.

We have explored other functional forms for modeling each cusp and found that the predictions for particle production and entanglement are not highly sensitive to the specific form of the function. Instead, the global quasi-periodic structure predominantly determines the magnitude of these effects. The function in \eqref{eq:OneCuspScatteringLength} strikes a good balance between accurately modeling experiments and maintaining simplicity.

\subsection{Covariance matrix in terms of Bogoliubov coefficients and the initial state}
In experiments such as the one described in section \ref{sec:Experiment}, the system of interest consists of linear perturbations of the condensate, which evolve during a finite time interval $t_{\rm f} - t_{\rm i}$ where the expansion/contraction phase takes place, and then are left to evolve freely. Since the condensate is homogeneous and isotropic at all times, each subsystem $(\vk, -\vk)$ evolves independently. Thus, assuming that each subsystem is prepared at $t_{\text{i}}$ in a Gaussian state with zero mean and covariance matrix $\bs{\sigma}_{\vk}^{\rm in}$, at $t_{\rm f}$ it will be in a Gaussian state with covariance matrix
\begin{equation}
    \bs{\sigma}_{\vk}^{\rm out}=\bS_\tk\cdot\bs{\sigma}_{\vk}^{\rm in}\cdot\bS_\tk^\top.
\label{eq:GaussianEvolution}
\end{equation}
Here $\bS_\tk$ is the evolution matrix given by $\eqref{eq:SmatrixTh}$ in terms of Bogoluibov coefficients.

Expression \eqref{eq:GaussianEvolution} allows us to predict observables in terms of the Bogoliubov coefficients and the details of the initial state. Recall that, as explained right after ~\eqref{eq:SmatrixTh}, the Bogoliubov coefficients can be readily computed by writing the OUT modes as linear combinations of IN modes.

We will use the parameterization of the covariance matrix as written in \eqref{eq:CovarianceMatrixOffAmpPhase}, in terms of amplitude, offset, and phase, since this will permit a direct comparison with experimental data.

For the initial state, we assume a thermal state at temperature $T$, for which $B^{\rm in}_\tk=\frac{1}{2} (1+2N^{\rm in}_\tk$), and $A^{\rm in}_\tk=\theta_\tk^{\rm in}=0$, where $N^{\rm in}_\tk = \left(\exp \left(\frac{\hbar \omi}{k_B T}\right) - 1\right)^{-1}$ is the mean number of thermal quanta in the mode $(\omega_\tk, \vk)$ present in the initial state. 

With this, it is straightforward to obtain the elements of the covariance matrix of the final state at $t>t_{\text{f}}$
\begin{equation}
\begin{split}
    A_\tk &= \left(1+2N_\tk^{\text{in}}\right) \abs{\alpha_\tk\beta_\tk}\,, \\
    B_\tk &= \left(1+2N_\tk^{\text{in}}\right)\left(\frac{1}{2} + \abs{\beta_\tk}^2\right)\, ,\\
    \theta_\tk &= \text{Arg}(\alpha_\tk\beta_\tk).
    \label{eq:OffAmpBogoliubov}
\end{split}
\end{equation}
The constrain $|\alpha_\tk|^2-|\beta_\tk|^2=1$ satisfied the Bogoluibov coefficients implies the following relation between amplitude and offset 
\begin{equation}
\begin{split} 
A_\tk &= \sqrt{B^2_\tk-\frac{\left(1+2N_\tk^{\text{in}}\right)^2}{4}}\,.
    \label{eq:AmpOfOff}
\end{split}
\end{equation}

From an experimental viewpoint, one can proceed either by (1) assuming the above relation holds and measuring only the offset or amplitude, as well as the temperature of the initial state, or by (2) measuring the offset and amplitude as independent variables and extracting the temperature of the initial state from the above relation (which can then be checked with an independent measurement). The second method only requires information about the state at late times, and it is the procedure followed in the experiment described in section \ref{sec:Experiment}, so we will stick to it.

\subsection{Entanglement}
\label{entanglement}

There exists a plethora of measures, witnesses, and quantifiers of bipartite entanglement, with varying levels of computational and experimental difficulty. In this article, we will work with two of them: the Logarithmic Negativity (LN) \cite{peres96, plenio05} and a Cauchy-Schwarz (CS) inequality discussed in \cite{PhysRevA.89.043808,Busch:2013gna,busch14}.

LN is defined from the density matrix of the bipartite system, $\hat{\rho}_{AB}$, as
\begin{equation}
    \LN(\hat{\rho}_{AB}) = \log_2 \|\hat{\rho}_{AB}^{\top_A}\|_1 \, ,
\end{equation}
where $\|\cdot\|_1$ denotes the trace norm and $\hat{\rho}_{AB}^{\top_A}$ is the 
\emph{partially transposed} density matrix with respect to subsystem $A$. A non-zero value of LN corresponds to the violation of the Positivity of Partial Transpose (PPT) criterion \cite{Peres:1996dw,Plenio:2005cwa}, which is a sufficient condition for entanglement. 

When restricted to Gaussian states, and if at least one of the subsystems is composed of a single mode, a non-zero value of LN is both necessary and sufficient for entanglement \cite{serafini17QCV}. Furthermore, LN becomes a quantifier for these simple systems, meaning that a larger value of LN implies a greater amount of entanglement. For pure states, LN grows monotonically with the entanglement entropy, which is the reference entanglement quantifier for pure states. However, in contrast to entropy, LN quantifies entanglement even when $\hat{\rho}_{AB}$ is a mixed state---which is the cases in any experimental set up at non-zero temperature. 

Moreover, for Gaussian states LN is easily computable in terms of the symplectic eigenvalues $\tilde{\nu}_I$ of the partially transposed covariance matrix\footnote{The partially transposed covariance matrix $\tilde{\sigma}_{AB}$ is obtained from $\sigma_{AB}$ by reversing the sign of all the components in $\sigma_{AB}$ involving momenta or, equivalently, by switching the rows and columns corresponding to the creation and annihilation operators of the subsystem with respect to which the partial transpose is performed. On the other hand, the symplectic eigenvalues of $\tilde{\sigma}_{AB}$ are defined as the absolute value of the eigenvalues of the product of matrices $\tilde{\sigma}_{AB}\cdot \Omega^{-1}$ \cite{serafini17QCV}. Although, for a system made of $N$ modes  $\tilde{\sigma}_{AB}\cdot \Omega^{-1}$ is a $2N\times 2N$ matrix, its eigenvalues appear in pairs of the form $\pm \ci \tilde \nu_I$, with $I=1,\cdots,N$ and $\tilde \nu_I\in \mathbb{R}$; thus, there are $N$ symplectic eigenvalues $\tilde \nu_I$. In this article, we study two modes at a time, hence $N=2$ in our calculations.} $\tilde{\sigma}_{AB}$ as \cite{serafini17QCV}
\begin{equation}
    \LN(\sigma_{AB}) = \sum_{I} \text{Max}\left[0, -\log_2 \tilde{\nu}_I\right]\,.
    \label{eq:LN}
\end{equation}
From this expression, it follows that subsystems $A$ and $B$ are entangled if and only if there is at least one symplectic eigenvalue $\tilde{\nu}_I$ smaller than $1$. In our case, the two symplectic eigenvalues of the partially transposed covariance matrix for each pair of modes $(\vk,-\vk)$  are $\tilde{\nu}_\tk=2\abs{B_\tk \pm A_\tk}$. Since $A_\tk \geq 0$ and $B_\tk \geq 1/2$, it follows that $2(A_\tk + B_\tk) \geq 1$. Therefore, there is only one symplectic eigenvalue of the partially transposed covariance matrix potentially smaller than one, namely 
\begin{equation}
    \tilde{\nu}_\tk^{\rm min} = 2(B_\tk - A_\tk)\,,
\label{eq:EigenvaluesOffAmp}
\end{equation}
where we have used that $A_\tk \leq B_\tk$. Thus, LN tells us that there is entanglement in our system if and only if $B_\tk - A_\tk < 1/2$.

On the other hand, for a bipartite system $(A,B)$ made of two single-mode subsystems, the CS inequality we will use reads
\begin{equation}
    \Delta = \abs{\langle \hat{a}_{A}\hat{a}_{B}\rangle}^2 - \langle\hat{a}_{A}^\dagger\hat{a}_{A}\rangle \langle\hat{a}_{B}^\dagger\hat{a}_{B}\rangle > 0.
    \label{eq:CS}
\end{equation}
This inequality provides, in general, a sufficient condition for entanglement. However, like LN, in the situation under consideration, $\Delta > 0$ is a necessary and sufficient condition for entanglement. This can be seen by noticing that, for our system, the inequality $\Delta > 0$ is equivalent to
\begin{equation}
    \Delta_{\tk} = A_\tk^2 - \left(B_\tk - 1/2\right)^2 \,.
    \label{eq:CSOffAmp}
\end{equation}
Since $A_\tk \geq 0$ and $B_\tk \geq 1/2$, this is equivalent to $B_\tk - A_\tk < 1/2$, which is the same condition we found for LN. Hence, we can use either $\text{LN} > 0$ or $\Delta > 0$ as faithful witnesses of entanglement.
%
\subsection{Effects of temperature and losses in entanglement production}\label{sec:ProdEnt}
This subsection is devoted to making quantitative predictions for the entanglement generated from pair creation between modes $\vk$ and $-\vk$, given a specific expansion history. We will further investigate how such entanglement varies with losses and ambient temperature, which are known to be the dominant channels of decoherence (see, e.g., \cite{Agullo:2022mvv}).

Using \eqref{eq:OffAmpBogoliubov}, we can write $\Delta_\tk$ and $\tilde\nu_{\tk}^{\rm min}$ in terms of Bogoliubov coefficients as
\begin{equation}
\begin{split}
    &\Delta_{\tk}=B_\tk-\frac{(1+2N_\tk^{\rm in})^2+1}{4}\,,
    \\
    &\tilde\nu_{\tk}^{\rm min}=2B_\tk-\sqrt{4 B^2_\tk-\left(1+2N_\tk^{\text{in}}\right)^2},
\end{split}
\label{eq:DeltaThBN}
\end{equation}
where 
$B_\tk= \left(1+2N_\tk^{\text{in}}\right)\left(\frac{1}{2} + \abs{\beta_\tk}^2\right)$
and $N^{\rm in}_\tk = \left(\exp \left(\frac{\hbar \omi}{k_B T}\right) - 1\right)^{-1}$.

The expression for $\tilde\nu_{\tk}^{\rm min}$ shows that it grows monotonically with $\abs{\beta_\tk}$ and decreases with $N_\tk^{\rm in}$, indicating that the logarithmic negativity (LN), and thus the entanglement in the final state, follows the same trend. Consequently, the experimental search for entanglement requires operating at the lowest possible temperatures to minimize thermal noise and enhance the entanglement.

\begin{figure}[t!]
    \centering
    \includegraphics[width=0.45\textwidth]{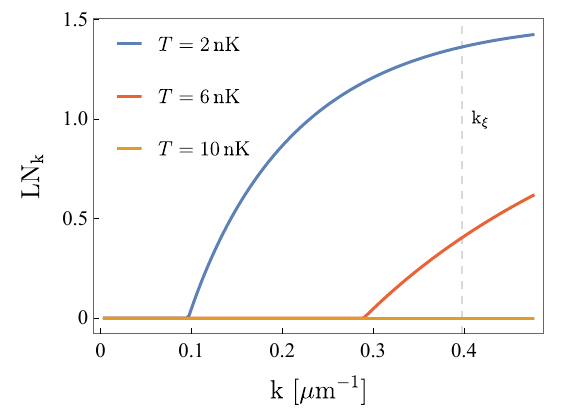}
    \caption{LN$_{\tk}$ versus wavenumber $k$, for $a(t) \propto t$ with a duration $t_{\rm f}-t_{\rm i} = 0.1 \, \text{ms}$. To produce this figure, we have used $\alpha_{s,\text{i}} = 400 \, a_B$ and $\alpha_{s,\text{f}} = 50 \, a_B$, which corresponds to approximately one $e$-fold of expansion, i.e., $\ln{a_{\text{f}}/a_{\text{i}}} \simeq 1.04$. The figure shows LN$_{\tk}$ for three different initial temperatures $T$. The gray dashed line denotes $\tk_{\xi}$ (for the most restrictive value of the scattering length); this value sets the boundary of the hydrodynamic regime. The plot indicates that temperatures $\lesssim \SI{10}{\nano\kelvin}$ are needed to produce entanglement within the hydrodynamic regime. This plot does not include losses, which would further degrade  entanglement.}
    \label{fig:LinearRampLN}
\end{figure}

To illustrate this, Figure~\ref{fig:LinearRampLN} shows predictions for $\text{LN}_{\tk}$ in a typical single-ramp experiment, as performed in \cite{Viermann:2022wgw}, in which the scale factor grows linearly with time. We observe that particle production is more pronounced for shorter duration and larger accumulated expansion (i.e., larger Hubble rates $\dot{a}/a$). In Figure~\ref{fig:LinearRampLN}, we choose favorable values of these parameters compared to those achieved in experiments (e.g., the lowest currently attained temperatures are \SI{10}{\nano\kelvin} \cite{NewPaper}).

The figure shows that, with this setup, entanglement is only produced, within the hydrodynamic regime $\tk < \tk_{\xi}$, for temperatures  $T \lesssim 10 \, \text{nK}$. Importantly, note that we have not included losses nor uncertainties in measurement in Fig.~\ref{fig:LinearRampLN}, which, as we will show below, further hinder the perspective of detecting entanglement with such a linear ramp $a(t)\sim t$.

These results advocate for experiments involving multiple expansion-contraction cycles, where entanglement production can be enhanced through resonances in pair production. This will be analyzed in the next subsection. 

On the other hand, to assess the effects of losses and detector inefficiencies we consider a simple nonetheless realistic loss model, in which a quantum is detected with probability $\eta$ and lost with probability $1-\eta$. 

Mathematically, this can be implemented by the transformation $\hat{a}_{\vk} \to \sqrt{\eta}\, \hat{a}_{\vk} + \sqrt{1-\eta}\,  \hat{e}_{\vk}$, where $\hat{e}_{\vk}$ represents the annihilation operator of an environment mode---assumed to be in a Gaussian state, so that the loss channel preserves Gaussianity. This transforms the covariance matrix of the acoustic perturbations as
\begin{equation}
\begin{split}
   \sigma_{\vk}^{\text{out}} &\to \eta\, \sigma_{\vk}^{\text{out}}+(1-\eta)\, \sigma_{\vk}^{\text{vac}}
\end{split}
\label{eq:LossModel}
\end{equation}
where $\sigma_{\vk}^{\text{vac}}$ is the covariance matrix of the vacuum.

If we parameterize the covariance matrix  by amplitude, offset and phase, as done in \eqref{eq:CovarianceMatrixOffAmpPhase}, this is equivalent to 
\begin{equation}
    \begin{split}
        B_k &\to \eta B_k + \frac{1-\eta}{2}, \\
        A_k &\to \eta A_k, \\
        \theta_k &\to \theta_k.
        \label{eq:OffAmpLosses}
    \end{split}
\end{equation}
Note that the phase $\theta_\tk$ is independent of losses (it is also independent of the initial temperature, as shown in \eqref{eq:OffAmpBogoliubov}), meaning it only depends on the expansion history. This is why the analysis of pair production done in \cite{Viermann:2022wgw} focused on this quantity. However, entanglement is independent of $\theta_\tk$; it originates from $A_\tk$ and $B_\tk$, and the consideration of losses and ambient thermal noise are of primary importance for these quantities.

We will use the efficiency $\eta \in [0,1]$ to parameterize losses. To quantify their impact on entanglement, let us focus on $\Tilde{\nu}_\tk^{\rm min}$---in terms of which $\text{LN}_\tk$ is computed, as shown in \eqref{eq:LN}. Losses modify $\Tilde{\nu}_\tk^{\rm min} = 2(B_\tk-A_\tk)$ to 
\begin{equation}
\Tilde{\nu}_\tk^{\rm min} = 2\eta(B_\tk-A_\tk)+(1-\eta),
\end{equation}
from which it can be seen that $\Tilde{\nu}_\tk^{\rm min}$ increases monotonically as $\eta$ decreases. This, in turn, causes $\text{LN}_\tk$ to decrease.

\begin{figure*}[t!]
    \centering
\includegraphics[width=0.9\textwidth]{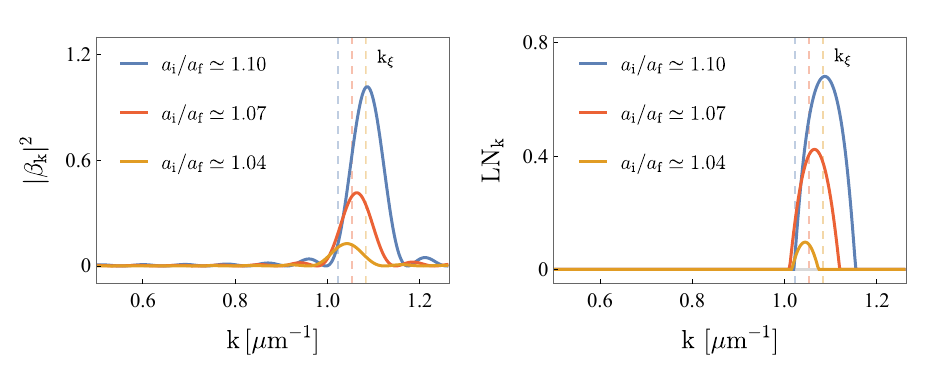}
    \caption{Number density of produced particles $\abs{\beta_\tk}^2$ (left panel) and LN$_{\tk}$ (right panel) versus wavenumber $\tk$, for three different values of the expansion/contraction accumulated within each cusp, $a_{\rm f}/a_{\rm i}$. This plot is obtained using $T = 12 \, \text{nK}$, $\alpha_{s, \text{f}} = 400 \, a_B$, $\Delta t_{\rm hold} = 0.75 \, \text{ms}$, $\delta = 0.4 \, \text{ms}$, $n = 12$, and $\eta_0 = 0.95$. The three curves correspond to $\alpha_{s, \text{i}}$ equal to $330$, $350$, and $370$ in units of the Bohr radius $\aB$, which  is equivalent to $a_{\rm i}/a_{\rm f} = 1.10$, $1.07$, and $1.04$, respectively. The vertical dashed line denotes the value of $\tk_{\xi}$ in each case, above which non-linear corrections to the dispersion (not included in this plot) become relevant. The figure shows the two effects discussed in the text: an increase in the peak values of $\abs{\beta_\tk}^2$ and LN$_{\tk}$ with $a_{\rm i}/a_{\rm f}$, and a shift in $º\tk_{\xi}$. For $a_{\rm i}/a_{\rm f}=1.10$, this shift causes the peak to fall entirely outside the hydrodynamical regime.}
\label{fig:EntanglementSpectraAlpha}
\end{figure*}

\subsection{Analysis of entanglement}

\begin{figure*}[t!]
    \centering
    \includegraphics[width=0.9\textwidth]{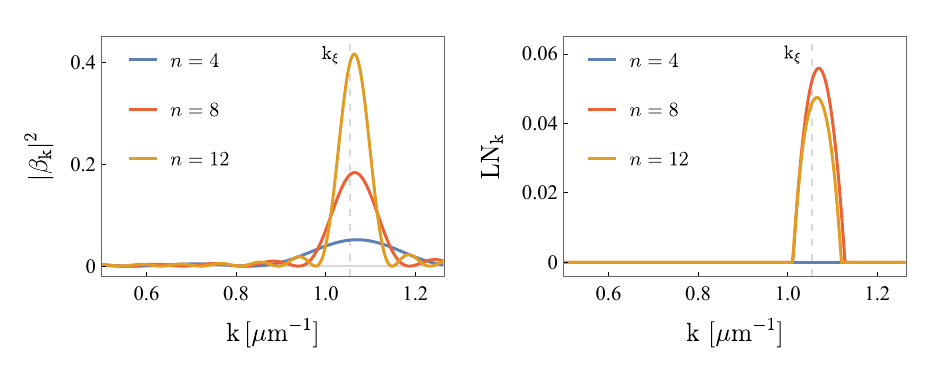}
    \caption{Number density of produced particles $\abs{\beta_\tk}^2$ (left panel) and LN$_{\tk}$ (right panel) versus wavenumber $\tk$, for three different numbers of cusps $n$. This plot is obtained using $T = 12 \, \text{nK}$, $\alpha_{s, \text{i}} = 350 \, a_B$, $\alpha_{s, \text{f}} = 400 \, a_B$, $\Delta t_{\rm hold} = 0.75 \, \text{ms}$, $\delta = 0.4 \, \text{ms}$, and $\eta_0 = 0.8$. The vertical gray dashed line denotes $\tk_{\xi}$, above which non-linear corrections to the dispersion (not included in this plot) become relevant. This figure shows the enhancement of peak values of $\abs{\beta_\tk}^2$ and LN$_{\tk}$ for moderate $n$, and the degrading effect losses have on entanglement for large $n$.}
\label{fig:EntanglementSpectraN}
\end{figure*}

\begin{figure*}[t!]
    \centering  \includegraphics[width=0.9\textwidth]{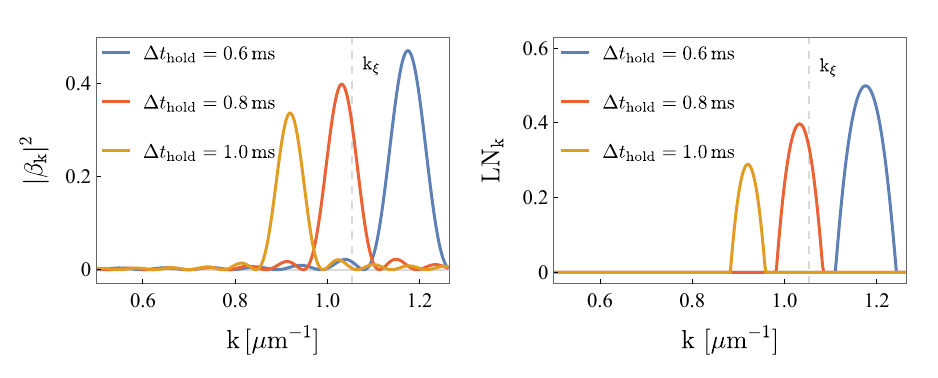}
    \caption{Particle number density $|\beta_{\tk}|^2$ (left panel) and LN$_{\tk}$ (right panel) for three different values of $\Delta t_{\rm hold}$ (see Figure~\ref{fig:ScatteringLength} for a graphical definition of this parameter). This plot is obtained using $T = 12 \, \text{nK}$, $\alpha_{s, \text{i}} = 350 \, a_B$, $\alpha_{s, \text{f}} = 400 \, a_B$, $\delta = 0.4 \, \text{ms}$ , $n = 12$ , and $\eta_0 = 0.95$. The vertical gray dashed line denotes $\tk_{\xi}$, above which non-linear corrections to the dispersion (not included in this plot) become relevant. We observe that $\Delta t_{\rm hold}$ controls the position of the resonance, shifting it toward the infrared as $\Delta t_{\rm hold}$ increases. It has also a an effects on the size of LN$_\tk$.}
    \label{fig:EntanglementSpectraThold}
\end{figure*}

\begin{figure*}[t!]
    \centering   \includegraphics[width=0.9\textwidth]{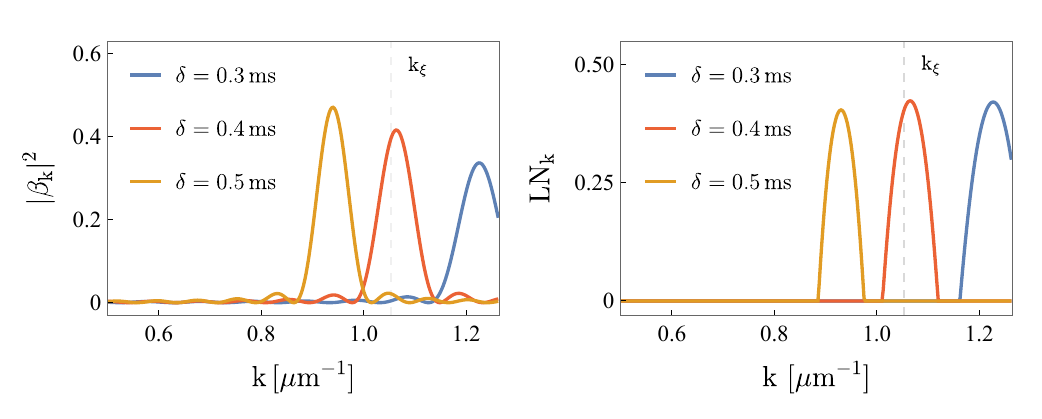}
    \caption{Particle number density $|\beta_{\tk}|^2$ (left panel) and LN$_{\tk}$ (right panel) versus $\tk$ for three different values of $\delta$ (see Figure~\ref{fig:ScatteringLength} for the meaning of this parameter; recall that a larger delta means broader and smoother cusps. This plot is obtained using $T = 12 \, \text{nK}$, $\alpha_{s, \text{i}} = 350 \, a_B$, $\alpha_{s, \text{f}} = 400 \, a_B$, $\Delta t_{\rm hold} = 0.75 \, \text{ms}$ , $n = 12$ , and $\eta_0 = 0.95$. The vertical gray dashed line denotes $\tk_{\xi}$, above which non-linear corrections to the dispersion (not included in this plot) become relevant. We observe that $\delta$ shifts the position of the resonance.}
\label{fig:EntanglementSpectraDelta}
\end{figure*}

Let us now focus on the expansion history shown in Figure~\ref{fig:ScatteringLength}, consisting of a periodic sequence of cusps. Four parameters control the details of this expansion history: (1) the parameter $\delta$, which determines both the ``sharpness'' and the duration of each cusp; (2) $\Delta t_{\rm hold}$, which sets the time interval between consecutive cusps; (3) the accumulated expansion/contraction at each cusp, $a_{\rm f}/a_{\rm i}$; and (4) the number $n$ of consecutive cusps.

Two more parameters control the environment and other aspects of the experiment, namely the ambient temperature $T$ and the efficiency $\eta$. The total efficiency depends on the duration of the experiment; keeping this in mind, we find it convenient to work with the {\em efficiency per cusp}, $\eta_0$. The total efficiency is then given by $\eta_0^n$.

Next, we discuss the impact that varying each of these parameters has on entanglement.

The effects of changing $\eta_0$ and $T$ have already been discussed above; specifically, higher temperatures and lower efficiencies degrade entanglement.
 
Regarding the accumulated expansion per cusp, $a_{\rm i}/a_{\rm f}$, we find that, as expected, increasing it enhances pair production and the generation of entanglement. However, changing $a_{\rm i}/a_{\rm f}$ also has a side effect: the wavenumber $\tk_\xi$, above which the hydrodynamic approximation breaks down, decreases as we increase $a_{\rm i}$---this is because increasing  $a_{\rm i}$ amounts to reducing the minimum value of the scattering length, $\alpha_{s,{\rm i}}$. This, in turn, reduces the range of wavevectors $\vk$ of interest, i.e., those within the hydrodynamic approximation. Our strategy is to fix a relatively high but attainable value of the scattering length at the peak of the cusp, $\alpha_{\text{s,f}} = 400 \, \aB$, and find the value of the scattering length at the valley of the ramp, $\alpha_{s,\rm{i}}$, that optimizes entanglement production within the hydrodynamic regime. Figure~\ref{fig:EntanglementSpectraAlpha} shows $\text{LN}_\tk$ for different accumulated expansions per cusp, highlighting the two effects just mentioned.

We also find that the value of $a_{\rm i}/a_{\rm f}$ which maximizes entanglement generation within the hydrodynamical regime depends slightly on the initial temperature. For the lowest attainable temperatures, it ranges around $a_{\rm i}/a_{\rm f}=1.07$, corresponding to $\alpha_{s,\text{f}} = 400 \, \aB$ and $\alpha_{s,\text{i}} = 350 \, \aB$. For these values, the hydrodynamical regime corresponds to $\tk \lesssim \tk_\xi \approx \SI{1.05}{\micro\meter}^{-1}$. 

Modifying the number $n$ of consecutive cusps also has a dual effect. On the one hand, the periodic variation of the scale factor imprints an oscillatory structure on the Bogoliubov coefficients $|\beta_\tk|$. We observe that the value of $|\beta_\tk|$ at the peaks increases with $n$. Thus, in an idealized experiment with no losses, a larger $n$ would lead to more entanglement. Notably,  entanglement  does not grow uniformly with $|\beta_\tk|$.  Instead, the rate of growth diminishes for large $|\beta_\tk|$, as it can be seen from \eqref{eq:DeltaThBN}.

However, increasing $n$ also extends the duration of the experiment, thereby increasing losses. As a result, there is an optimal number of cycles, $n_{\rm op}$, beyond which the degradation due to losses outweighs the enhancement of entanglement. The value of $n_{\rm op}$ depends on the efficiency per cusp, $\eta_0$, which must be characterized experimentally---a possible approach to achieve this is to measure the spectrum of produced quanta for identically prepared realizations that run over different numbers of cusps. Figure~\ref{fig:EntanglementSpectraN} illustrates the two effects that arise from modifying $n$.

The time lapse $\Delta t_{\rm hold}$ between cusps controls the periodic structure of the expansion history $a(t)$. It is therefore expected that $\Delta t_{\rm hold}$ determines the location of the resonant peaks observed in the coefficients $|\beta_\tk|$. This is indeed the case, as shown in Figure~\ref{fig:EntanglementSpectraThold}. This figure shows that increasing  $\Delta t_{\rm hold}$ shifts the resonances towards infrared wavenumbers $\vk$. One takeaway from this figure is that it is important to tune $\Delta t_{\rm hold}$ for the resonances in entanglement to fall within the hydrodynamical regime.

Finally, $\delta$ has also an effect in the position of the resonances. This is because this parameter controls the duration of the cusps (see Fig.~\ref{fig:EntanglementSpectraDelta}).

\section{Optimizing detectability of entanglement}\label{sec:Optimization}

The goal of this section is to present an exploration of the parameter space of the experiment and an identification of the optimal configurations for observing of the entanglement generated in the pair production process. 

We will argue that detection of entanglement with a statistical significance of  $\approx2\sigma$ can be reached within current experimental capabilities, and detection at 3-4$\sigma$ is achievable with either moderate improvements in the precision with which offset and amplitude are measured, or by lowering the temperature of the initial state from the current $12$nK to $8$nK.

We will denote by $\epsilon_{\text{r}}$ the relative error in the experimental reconstruction of amplitude 
$A_\tk$ and and offset $B_\tk$  from the measurement of density contrast correlations. This error propagates to an absolute error  in the value of the entanglement witnesses. Let us focus on one of them, for instance, the CS inequality  $\Delta_\tk$, and denote by $\bar \Delta_\tk$ and $\epsilon_{\Delta_\tk}$ the experimentally obtained mean and error of $\Delta_\tk$. We are interested in computing the mean value $\bar \Delta_\tk$ in units of the error $\epsilon_{\Delta_\tk}$. We will denote this quantity by $X_\tk$, and use it quantify how many errors $\epsilon_{\Delta_\tk}$ the measured value is away from zero. Once $X_\tk$ is computed, we will say entanglement can be detected with a significance of $X_\tk$ ``sigmas''. 

Our goal is to find the region in the parameter space where $X_{\vk}$ attains its maximum value. As a remainder, the free parameters in the analysis are $T$, $\eta_0$, $a_{\rm i}/a_{\rm f}$, $\delta$, $\Delta t_{\rm hold}$ and the norm of the wavenumber $\tk$. Furthermore, we will parameterize $a_{\rm i}/a_{\rm f}$ by the minimum value of the scattering length, $\alpha_{s,\rm i}$, with its maximum value kept fixed and equal to $\alpha_{s,\rm f}=400 \, \aB$. 

The optimization procedure goes as follows. We choose a value of $T$, $\eta_0$ and $\delta$, and compute $X_{\vk}$ scanning over the other four parameters. Namely, we use $\alpha_{s,\text{i}}\in[260\aB, 380\aB]$, with step $\SI{10}{\micro\meter}$; $\Delta t_{\rm hold}\in[\SI{0}{\milli\second},\SI{1.5}{\milli\second}]$, with step $\SI{0.1}{\milli\second}$; $n\in[1,12]$ in unit steps; and $\delta \in [0.1 \, \text{ms}, 1.0 \, \text{ms}]$ with step $0.1 \, \text{ms}$. For $T$ and $\eta_0$ we use a few realistic values, namely $T\in\{8,12,16,20\}$nK and $\eta_0\in\{0.98,0.95,0.9,0.85\}$ ---these values correspond to total losses around 20-85\% for the optimal number of cycles in each case. In experiments such as the one described in section \ref{sec:Experiment}, we expect lower total losses. Hence, our results fall in the conservative side. Finally, we explore a relative error in the determination of $A_\tk$ and $B_\tk$ in the range of 1\% to 6\% (relative errors with current capabilities are of 5\% \cite{NewPaper}).

\begin{figure*}[t!]
    \centering \includegraphics[width=0.8\textwidth]{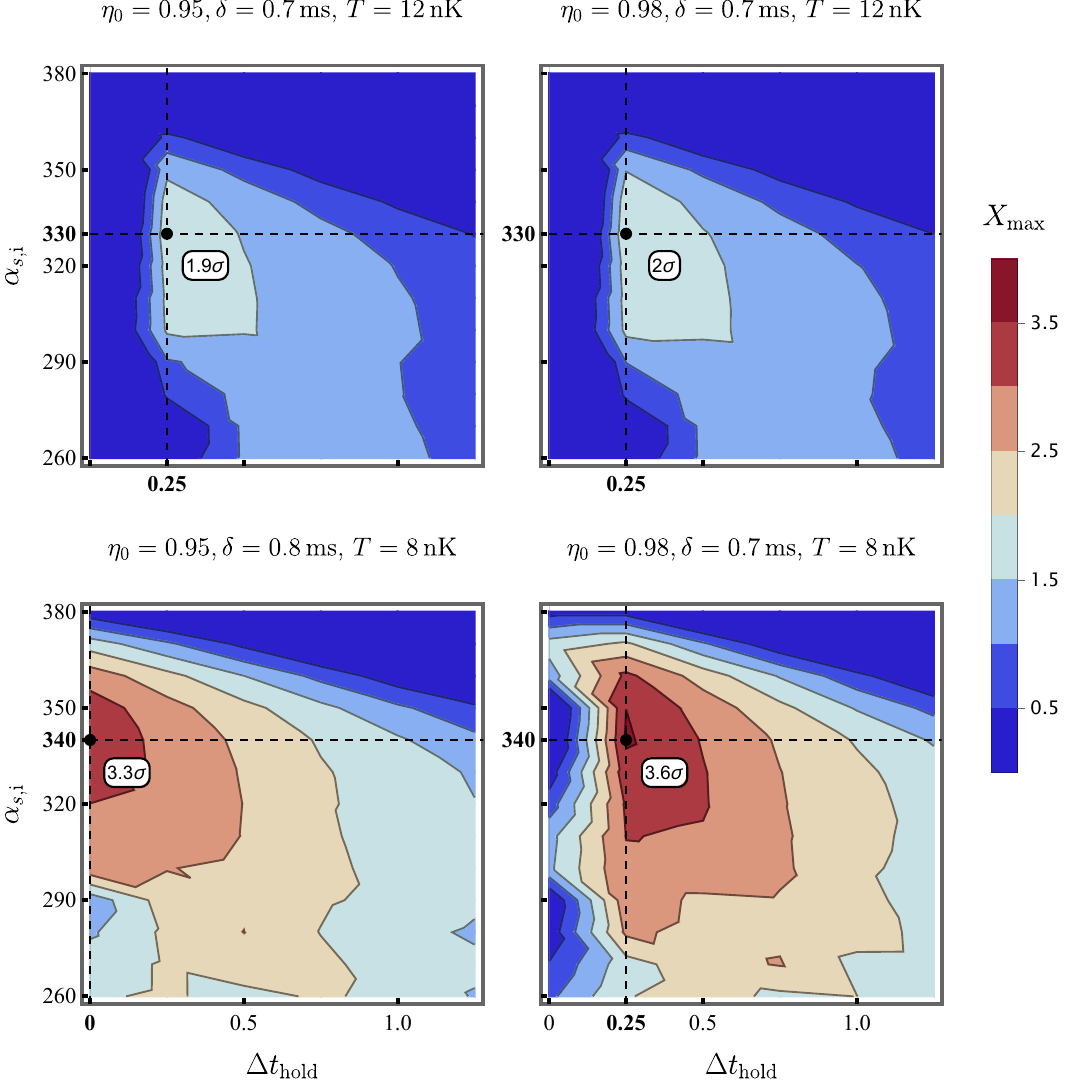}
    \caption{Contour plots in experimental parameter space of the maximum significance $X_{\rm max}$ of measuring entanglement production due to pair creation  in 2D BEC analog bouncing universe, within the hydrodynamic regime. The axis denote initial scattering length $\alpha_{s,\rm i}$ and waiting time between cusps $\tho$. The optimal configuration  for each of the considered initial temperatures $T$ and loss-per-cycle $\eta_0$ is marked with a black dot. For this plot, we have fixed $n=8$ in the optimization process. The value of $\tk$ which maximizes $X_{\rm max}$ (black dots), and the value of $\tk_{\xi}$ in the corresponding experimental configuration are $\tk_{\text{max}}=1.020 \, \mu\text{m}^{-1}, \tk_{\xi}=1.023\, \mu\text{m}^{-1}$ in the upper panels, and $\tk_{\text{max}}=1.020 \, \mu\text{m}, \tk_{\xi}=1.038\, \mu\text{m}$ in the lower panels. Finally, we have assumed a relative error of  5\% in the measurement of $A_\tk$ and $B_\tk$.}
    \label{fig:DetectabilityContours}
\end{figure*}

\begin{figure*}[t!]
    \centering   \includegraphics[width=0.8\textwidth]{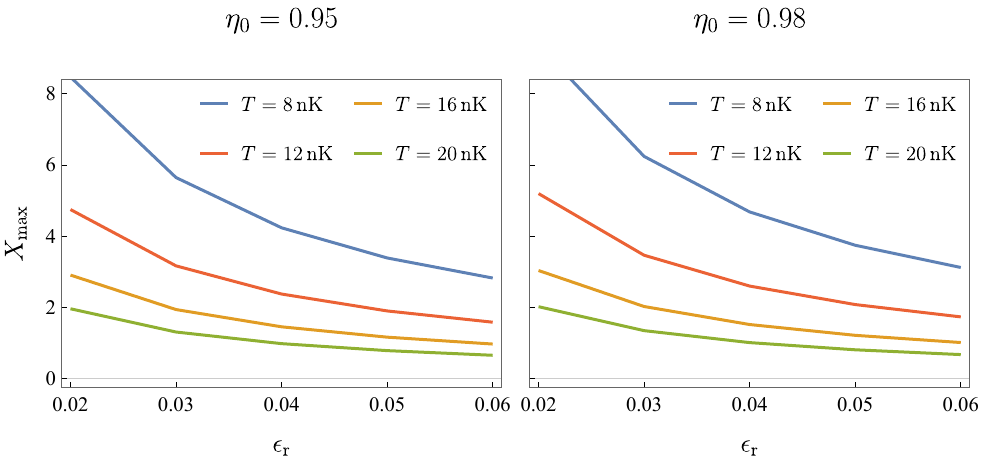}
    \caption{Significance at which entanglement can be detected, as a function of relative error $\epsilon_{\text{r}}$ in the experimental determination of $A_\tk$ and $B_\tk$ for optimal parameters $\alpha_{s,\text{i}}$, $n$, and $\tho$. The same precision is assumed to be achievable for both $A_\tk$ and $B_\tk$. This figure shows that, with currently available precision of $5-6\%$ and initial temperatures of $\sim12\,$nK, entanglement can be detected at around $2\sigma$. With some improvements in precision or temperature  the statistical significance rises up to $\sim 4\sigma$.}
    \label{fig:MaxSigmas}
\end{figure*}

For each value of these parameters, we maximize  $X_{\vk}$ varying the wavenumber within the interval $\tk\in[\SI{0.01}{\micro\meter},\tk_\xi]$ with step $\SI{0.01}{\micro\meter}$. The result, denoted as  $X_{\rm max}$ quantifies the detectability of entanglement in the considered set up. A representative sample of our  results are shown in Figure~\ref{fig:DetectabilityContours}.%
Furthermore, Figure~\ref{fig:MaxSigmas} shows dependence of $X_{\rm max}$ for the optimal configurations with the precision at which $A_\tk$ and $B_\tk$ are measured (we assume both are measured with the same precision, although this assumption can be easily relaxed). The figure shows that a modest improvement in precision and ambient temperature from currently attainable values ($5\%$ and $12\,$nK respectively \cite{PrivComm}) can increase the significance of the detection of entanglement up to $\sim4\,\sigma$. 

\begin{figure*}[t!]
    \centering  \includegraphics[width=0.8\textwidth]{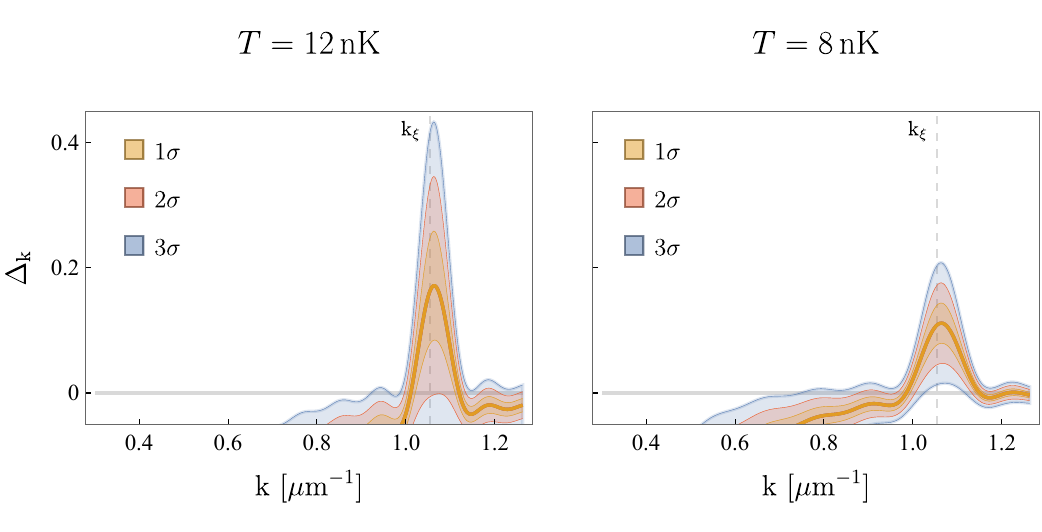}
    \caption{$\Delta_\tk$ versus $\tk$. The bands indicate $1\sigma$ (orange), $2\sigma$ (red) and $3\sigma$ (blue) contours, using a relative error for $A_\tk$ and $B_\tk$ of 5\%. This Figure uses $\eta_0=0.95$, $\alpha_{\rm s}=350 \, \aB, \tho=0.75 \, \text{ms}, \delta=0.4 \, \text{ms}$, and $n=12$ for the left panel as well as $n=9$ for the right panel, which are the optimal values. The vertical gray dashed line denotes $\tk_{\xi}$, above which non-linear corrections to the dispersion (not included in this plot) become relevant.}
    \label{fig:EntanglementSpectrum}
\end{figure*}

Finally, Fig.~\ref{fig:EntanglementSpectrum} shows the  $\Delta_\tk$ for the values of experimental parameters that optimize detection at the assumed temperature (12 and 8 nK) and precision (5\%). We observe that detectability of entanglement is best near the limits of the hydrodynamical regime, $\tk\approx \tk_{\xi}$. 

\begin{figure*}[t!]
    \centering  \includegraphics[width=0.8\textwidth]{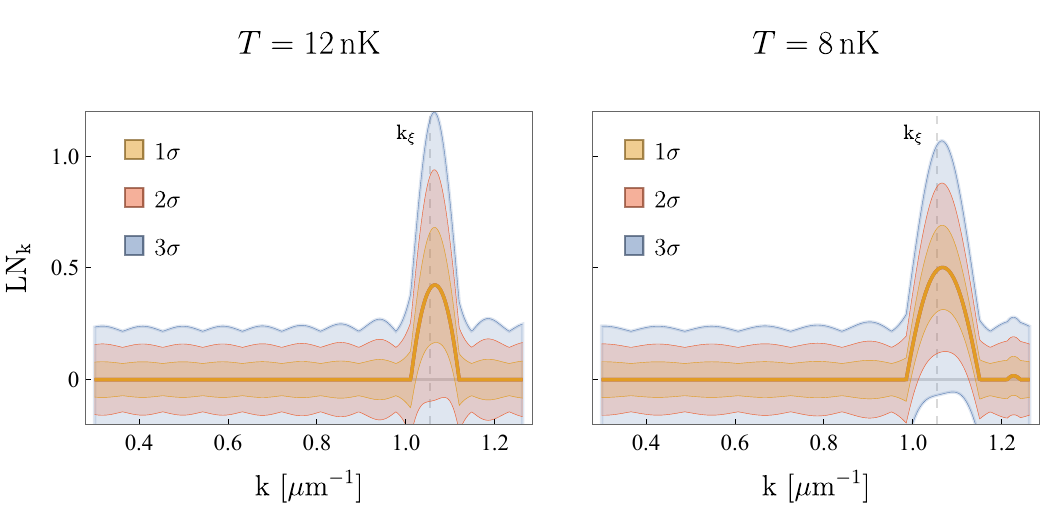}
    \caption{$\text{LN}_\tk$ versus $\tk$. The bands indicate $1\sigma$ (orange), $2\sigma$ (red) and $3\sigma$ (blue) contours, using a relative error for $A_\tk$ and $B_\tk$ of 5\%. This Figure~ uses $\eta_0=0.95$, $(\alpha_{\rm s}=350 \, \aB,\tho=0.75 \, \text{ms}, \delta=0.4 \, \text{ms})$, and $n=12$ for the left panel as well as $n=9$ for the right panel, which are the optimal values. The vertical gray dashed line denotes $\tk_{\xi}$, above which non-linear corrections to the dispersion (not included in this plot) become relevant.}
    \label{fig:EntanglementSpectrumLN}
\end{figure*}

For completeness, we also present the entanglement quantifier $\text{LN}_\tk$ in Fig.~\ref{fig:EntanglementSpectrumLN}. It is worth noting that $\Delta_{\tk}$ achieves slightly greater statistical significance in witnessing entanglement compared to $\text{LN}_\tk$\footnote{Since $\text{LN}_\tk$ and $\Delta_\tk$ are different functions of amplitude and offset, standard error propagation leads to a small difference in statistical significance of detection. This difference does not significantly alter our  results.}. This is why we have concentrated on $\Delta_{\tk}$ in this section.

As a last remark, let us note that our results depend on the definition of hydrodynamical regime $\tk<\tk_\xi$. Our definition of $\tk_\xi$ (the value of $\tk$ for which non-linear correction to the dispersion relations are $10\%$) is slightly more conservative than $\tk_\xi=\xi^{-1}$, the value commonly assumed to interpret experimental results. One could further restrict the range of wavenumbers $\tk$ to values for which the contributions of non-linear terms in the dispersion relations are even smaller than those we have already considered. Such restrictions, however, make the detection of entanglement more challenging. For instance, by restricting the range of $\tk$ to values where non-linear contributions in the dispersion relations are below $5\%$, the temperature needed to detect entanglement at $3\sigma$'s decreases to approximately $\SI{6}{\nano\kelvin}$. Although small, this temperature is expected to be within reach of near-term experiments.

\section{Outlook}\label{sec:Outlook}

Spontaneous particle creation in an expanding universe is just one instance of pair creation in the presence of non-trivial backgrounds, a fundamental phenomenon predicted by quantum field theory. Classical signals associated with this process have been observed \cite{eckel_rapidly_2018, Banik:2021xjn, Viermann:2022wgw,philbin_fiber-optical_2008, weinfurtner_measurement_2011, euve_observation_2016, munoz_de_nova_observation_2019, drori_observation_2019, shi_quantum_2023,torres_rotational_2017, braidotti_measurement_2022}. The next step is to obtain a unequivocal confirmation of its quantum origin 
(see \cite{Steinhauer2016,Leonhardt:2016qdi,Steinhauer:2018qzg} for work in this direction in analogue spacetimes, and \cite{Chen:2021xhd} for a detection of entanglement in a scenario with no spacetime analogue). The results of this article suggest that detecting entanglement from quantum pair creation with confident statistical significance is within reach. This would provide experimental confirmation of a longstanding non-perturbative prediction of QFT, in a regime where experimental tests of our most fundamental theories are scarce. Furthermore, positive results could pave the way for detecting entanglement in other phenomena predicted by quantum field theory in curved spacetimes.

In this article, we have analyzed the generation of entanglement in a two-dimensional BEC-based quantum simulator, theoretically and experimentally studied in \cite{CosmologyPaper2022,Viermann:2022wgw,Tolosa-Simeon:2022umw}. Our focus on this experimental platform stems from the observation that the setup described in \cite{Viermann:2022wgw} is on the edge of detecting entanglement generated by cosmological pair production.

We have concentrated our analysis on the hydrodynamic regime of perturbations of the BEC, where the analogy with quantum fields propagating in a curved spacetime is robust. While the analysis of entanglement in the dispersive regime $\tk > \tk_{\xi}$ is certainly of interest, such effects break the analogy with relativistic quantum fields propagating in curved FLRW spacetimes. We defer the study of these effects to future work.

Our results demonstrate that, with currently available techniques, it is possible to detect entanglement with a significance of $\approx 2\sigma$ within the hydrodynamic regime. Furthermore, modest improvements in the temperature of the system or measurement precision could increase the significance beyond $3.3\sigma$.

Our approach is conservative in the sense that it is grounded in experimental capabilities already demonstrated in \cite{Viermann:2022wgw}, along with
modest upgrades attainable in the short term. 

The most speculative aspect of our work lies in the model of losses we employ, which assumes that, to a good approximation, the losses do not alter the Gaussian nature of the phononic perturbation state. Although this is a reasonably mild assumption, it is crucial to thoroughly characterize experimental losses and detector inefficiencies when applying the tools developed in this paper to specific setups, as these factors can significantly impact entanglement detectability. Nonetheless, our analysis suggests that detecting entanglement remains feasible even with moderate losses.

To enhance the experimental detection of entanglement, we propose implementing a sequence of expansion-contraction cycles, rather than simulating a linear scale factor $a(t) \propto t$ and trying to maximize the number of accumulated $e$-folds of expansion. Expansion-contraction cycles induce resonances in particle production. These resonances can be experimentally tuned to boost the amount of entanglement produced, offering a clear advantage in terms of detectability. Interestingly, a similar idea was foreshadowed by Schr\"odinger in 1939 \cite{SCHRODINGER1939899}, when he highlighted the importance of ``critical periods of cosmology, when expansion changes to contraction and vice-versa'' in regards of particle production.

Future work could involve replacing the initial vacuum state with a single-mode squeezed state. As discussed in \cite{Agullo:2021vwj,Brady:2022ffk}, initial squeezing can stimulate further entanglement generation, countering the detrimental effects of temperature and losses, thus enhancing detectability.

Further extensions could also explore analog universes with spatial curvature \cite{CosmologyPaper2022}, or scale factors mimicking quasi-de Sitter expansions. Such explorations would allow us to investigate how spatial and spacetime curvature affects the distribution of entanglement across different modes. Considerable theoretical efforts are already underway to understand the structure of entanglement in real space within quantum field theory \cite{Bianchi:2019pvv,Martin:2021xml,Martin:2021qkg,Espinosa-Portales:2022yok,Agullo:2022ttg,Agullo:2024har,Agullo:2024har}, particularly in spacetimes with various geometric properties \cite{Brahma:2023uab,Brahma:2023lqm,K:2023oon,Ribes-Metidieri:2024vjn,Ribes-Metidieri:2024vjn}. The experimental platforms discussed here are well suited to test the structure of quantum entanglement in field theories and its interplay with the curvature of spacetime. 

\begin{acknowledgements}
We thank Markus Oberthaler's group, and specially Elinor Kath, Marius Sparn, and Nikolas Liebster
for discussions on details of the experimental setup. We also thank Elinor Kath, Marius Sparn, and Victor Gondret for their detailed feedback on a preliminary version of the manuscript. APL also thanks Stefan Floerchinger's group for discussions on the topic.
IA and AD are supported by the Hearne Institute for Theoretical Physics, by the NSF grants PHY-2409402, PHY-2110273, PHY-1903799, and PHY-
220655, and by the RCS program of Louisiana Boards of Regents through the grant LEQSF(2023-25)-RD-A-04. IA is also supported by the Visiting Fellow program of Perimeter Institute for Theoretical. Research at Perimeter Institute is supported in part by the Government of Canada through the Department of Innovation, Science and Industry Canada and by the Province of Ontario through the Ministry of Colleges and Universities. APL acknowledges support through the MICINN (Ministerio de Ciencia, Innovación y Universidades, Spain) fellowship FPU20/05603 and project PID2020-118159GBC44, and COST (European Cooperation in Science and Technology) Actions CA21106 and CA21136. AD and APL acknowledge support from the project PID2022-139841NB-I00. AD also acknowledges support to Grant PR28/23 ATR2023-145735 funded by
MCIN/AEI/10.13039/501100011033. APL also thanks the Hearne Institute for Theoretical Physics at Louisiana State University for its hospitality during the course of this investigation.
\end{acknowledgements}


\bibliography{references.bib}

\end{document}